\documentclass[11pt]{article}
\pdfoutput=1 

\usepackage{jheppub} 

\usepackage{stackrel}
\usepackage[T1]{fontenc} 
\usepackage[title]{appendix}
\usepackage{booktabs}
\usepackage{amsmath}

\makeatother

\begin{document}

	\vspace*{4mm}
	
	\thispagestyle{empty}

	\begin{center}
		
			\begin{minipage}{20cm}
				\begin{center}
					\hspace{-5cm }
					\LARGE
					\sc
					Anomaly induced transport
					\\
					\hspace{-5cm} in holographic superconductors
				\end{center}
			\end{minipage}
			\\[15mm]
			
			\renewcommand{\thefootnote}{\fnsymbol{footnote}}

			{\large
				Ashis Tamang$^{\, a}$ \footnote{ashismaniram@gmail.com},
				Nishal Rai$^{\, b}$ \footnote{nishalrai10@gmail.com},
				Javier Esteban$^{\, c}$ \footnote{javiereq@correo.ugr.es},
				Eugenio Meg\'{\i}as$^{\, c}$ \footnote{emegias@ugr.es} \,
			}\\[12mm]
		\end{center}
		\noindent

		${}^{a}\!$
		\textit{Department of Physics, SRM University Sikkim, Upper Tadong, Sikkim, 737102,  India}

		${}^{b}\!$
		\textit{Department of Physics, St. Joseph's College,  Darjeeling, North Point, 734104, India}

		${}^{c}\!$
		\textit{Departamento de F{\'\i}sica At\'omica, Molecular y Nuclear and} \\
		\indent \; \textit{Instituto Carlos I de F{\'\i}sica Te\'orica y Computacional,}  \\
		\indent \; \textit{Universidad de Granada, Avenida de Fuente Nueva s/n, E-18071 Granada, Spain}

		\vspace*{4mm}

		\begin{center}
			{  \bf  Abstract }
		\end{center}
		\begin{minipage}{15cm}
  We investigate the anomaly-induced transport properties of a holographic superconductor exhibiting spontaneous symmetry breaking. The holographic model incorporates vector and axial gauge fields, an additional non-anomalous $U(1)$ symmetry, and a charged scalar field that drives the phase transition. Working in the probe limit, we numerically construct the superconducting background under two alternative scalar field quantizations, corresponding to operators with scaling dimensions $\Delta=1$ and $\Delta=3$ (denoted as the $O_1$ and $O_2$ theories, respectively). Using Kubo formulae, we compute the chiral magnetic conductivities and examine their response to the scalar condensate. Our findings demonstrate that spontaneous symmetry breaking allows anomaly-induced transport to extend beyond anomalous currents, significantly impacting the non-anomalous sector as well. Remarkably, this effect appears in both theories, indicating that the influence on the non-anomalous sector is independent of the chosen quantization. The comparison of our numerical results with the analytical limits for $O_i \to 0$ and $O_i \to \infty$ $(i=1,2)$ reveals consistent behavior across both theories.
			\vspace{0.5cm}
		\end{minipage}

	      \newpage
		\tableofcontents

\setcounter{footnote}{0}
		
		\section{Introduction}
		Quantum anomalies arise from a conflict between two of the most sacred principles of theoretical physics: \textit{symmetry} and \textit{quantum mechanics}. On the one hand, symmetry gives us conservation laws and selection rules, while on the other hand, quantum mechanics tells us that particles can fluctuate in ways that can break this symmetry at the quantum level. The well-known example of this phenomenon is the decay of neutral pions into two photons. While this process is classically forbidden, it actually occurs in nature due to the presence of the axial anomaly \cite{Bell:1969ts,Adler:1969gk}. Though the decay of the pion illustrates the microscopic manifestation of the axial anomaly, its macroscopic outcomes are equally profound. In a system that exhibits chiral imbalance, such as a relativistic fluid, quantum anomalies give rise to dissipationless transport phenomena \cite{Son:2009tf}. The most celebrated examples are the chiral magnetic effect (CME), where an electric current is induced parallel to an external magnetic field \cite{Fukushima:2008xe}, and the chiral vortical effect (CVE), where an electric current is generated parallel to the fluid vorticity \cite{Landsteiner:2011cp}. These are examples of anomaly-induced transport phenomena, meaning that quantum corrections to classical symmetries are ultimately responsible for the generation of non-dissipative currents, representing fascinating manifestations of quantum anomalies in macroscopic systems.
		
		Anomaly-induced transport has played a significant role across multiple domains of physics. In high-energy physics, the search for signatures of the CME and CVE in heavy-ion collisions is an ongoing endeavor, motivated by the possibility of observing event-by-event parity violation in the Quark-Gluon Plasma, and has culminated in the isobar program at RHIC and the LHC \cite{STAR:2009wot,Kharzeev:2019zgg}. In astrophysics, anomalous transport has been suggested to explain neutron star kicks \cite{Kaminski:2014jda} and the origin of primordial magnetic fields in cosmology \cite{Giovannini:1997eg}. In condensed matter physics, the CME is responsible for the large enhancement of longitudinal conductivity observed in Weyl and Dirac semimetals subjected to a magnetic field \cite{Li:2014bha,Gooth:2017mbd}, while quantum Hall physics is understood via anomaly inflow from the bulk to the boundary of topological insulators \cite{Tong:2016kpv,Callan:1984sa}. These examples highlight the ubiquity of anomaly-induced transport across energy scales, bridging fundamental physics and observable phenomena.
		
		A powerful tool for studying these strongly coupled systems is the AdS/CFT correspondence often referred to as the gauge/gravity duality or holography $-$ which states that a gravitational theory in a ($d+1$)$-$dimensional bulk spacetime is dual to a quantum field theory living on its $d$$-$dimensional boundary. Perhaps the most widely known example of this correspondence is the duality between ${\mathcal N} = 4$ supersymmetric Yang-Mills (SYM) theory in the large $N$ limit and type IIB string theory in the AdS$_5$ $\times$ S$^5$ background, proposed by Juan Maldacena in 1997 \cite{Maldacena:1997re,Witten:1998qj,Gubser:1998bc}.  Within this framework, quantum anomalies in the field theory are encoded through Chern-Simons terms in the bulk, with a pure gauge Chern-Simons term corresponding to the axial anomaly~\cite{Kharzeev:2009pj}, and a mixed Chern-Simons term to the mixed gauge-gravitational anomaly~\cite{Landsteiner:2011cp}, thus providing a systematic way to compute transport coefficients by using different methods.

		A central question in this context  is whether anomaly-induced transport is restricted to the anomalous currents or it can be extended to the non-anomalous sector in the presence of symmetry breaking. In our recent work \cite{Tamang:2025ekj}, we addressed this question by using a holographic model with a pure gauge and mixed gauge-gravitational Chern-Simons terms, and found that indeed quantum anomalies affect the non-anomalous currents when symmetries are explicitly broken, confirming that the effects of quantum anomalies are not confined to their original source. Now the question is: \textit{what would happen if the symmetry is spontaneously broken?  Will this phenomenon still persist?} This is precisely what we will explore in the present work.
		
		To address this question, we use the framework of a holographic superconductor in a model including a charged scalar. Above a critical temperature $T_c$, the scalar field is zero and no condensate forms; this is called the normal phase. As we cool the system below $T_c$, the scalar field condenses: this corresponds to the formation of a condensate in the dual field theory. In this situation the symmetry is spontaneously broken, and the system transitions into a holographic superconductor~\cite{Hartnoll:2008kx,Hartnoll:2008vx,Amado:2009ts,Musso:2013vtg,Cai:2015cya}. In this setup, we will work in the probe limit so that metric fluctuations can be neglected and include only the pure gauge Chern-Simons term in the action of the model, which encodes the axial anomaly. More specifically, we will consider a model with three symmetries, two of which can be identified with the usual vector $(V)$ and axial $(A)$ $U(1)$ symmetries of Dirac fermions. Of these, only the axial symmetry is anomalous. In addition, we consider a non-anomalous symmetry $(W)$,~\footnote{To be precise, we note that the vector symmetry is also non-anomalous. The vector gauge field does however appear in the Chern-Simons term, which means that it has a {\em covariant} but not a {\em consistent} anomaly.} and a scalar field~$(\phi)$ that is simultaneously charged under the axial and the non-anomalous symmetry through the covariant derivative $D_M \phi =\left[\partial_M-i(A_M-W_M)\right]\phi$.  Among other transport properties, we will study whether the axial anomaly generates an induced current associated to the non-anomalous symmetry, i.e. $\vec{J}_w \ne 0$, in the presence of spontaneous symmetry breaking (SSB). The scalar field admits two possible quantizations \cite{Horowitz:2008bn}, corresponding to operators of dimension $\Delta=1$ and $\Delta=3$. This allows us to study transport phenomena in two different quantization schemes, providing a controlled framework to compare the conductivities and to assess whether the effect of anomaly-induced transport into the non-anomalous sector is universal or depends on the specific choice of the dual operator. In the present work, the anomalous conductivities will be computed by using Kubo formulae.
		
		The structure of this manuscript is as follows. In section ~\ref{sec:action}, we introduce the holographic model, which incorporates vector and axial gauge fields, an additional non-anomalous gauge field, and a charged scalar that drives spontaneous symmetry breaking. We then define the ansatz for the background fields and derive the equations of motion within the probe limit, solving them numerically by using the shooting method. With the chosen value for the scalar mass $m$, the model yields two theories, $O_1$ and $O_2$ theories, corresponding to operators of dimensions $\Delta=1$ and $\Delta=3$, respectively. We determine the background solutions for both cases. Subsequently, in section~\ref{sec:flut}, we examine the transverse fluctuations of the gauge fields atop the superconducting backgrounds established in section~\ref{sec:action}. In section~\ref{sec:kubo_formulae}, we outline the constitutive relations for the currents and present the Kubo formulae required to compute the chiral magnetic conductivities. In section \ref{sec:result}, we present the main results for the transport coefficients. We begin by describing in subsection \ref{subsec:numerical_method} the numerical method used to solve the coupled equations of motion for the fluctuations. In subsections \ref{sec:result1} and \ref{sec:result2}, we present the conductivities for the $O_1$ and $O_2$ theories, respectively, as functions of their corresponding condensates. In order to provide a consistency check of the numerical computation, we also compare our numerical results against some analytic limits that will be determined later. Additionally, in subsection~\ref{subsec:disc_vv_aa_ww}, we characterize the functional dependence of the conductivities on the condensates and chemical potentials. Furthermore, for completeness, we examine the conductivities at vanishing chemical potentials in subsection~\ref{sec:result0}. We summarize our main findings and discuss possible future directions in section~\ref{sec:discussion}. Finally, appendix~\ref{sec:appendix_EoM_fluctuations} contains the explicit equations of motion for the fluctuations, we derive in appendix~\ref{sec:sigma_analytic} the analytical results for the conductivities in the $O_i \to 0$ and $O_i \to \infty$ limits (for $i=1,2$) across different bases, and appendix~\ref{sec:appendix_sigma} collects additional numerical results for the conductivities as a function of~$T/T_c$.

\section{Holographic action}
\label{sec:action}
		
		In this work, we consider a holographic model that includes a pure gauge Chern-Simons terms in the action which mimics the axial anomaly. The action reads \cite{Landsteiner:2013aba,Megias:2016ery}
		\begin{eqnarray}
			\mathcal S &=& \dfrac{1}{16 \pi G} \int d^5x\sqrt{-g}
			\Big[
			R + 2\Lambda -\dfrac{1}{4}  F_V^{2} \nonumber -\dfrac{1}{4}  F_A^{2} \nonumber -\dfrac{1}{4}  F_W^{2} \nonumber \nonumber \\&&+ \epsilon^{MNPQR} A_M \left(\dfrac{\kappa}{3}(F_A)_{NP}(F_A)_{QR}+{\kappa}(F_V)_{NP}(F_V)_{QR} \right)\Big] +  \mathcal S_{\rm GH}   \,,  \label{act}
		\end{eqnarray}
		where $\Lambda = 6 / \ell^2$  is the cosmological constant with $\ell$ the radius of AdS. The field-strength tensors are defined as $(F_S)_{MN}=\partial_M S_N-\partial_N S_M$ for $S = V, A, W$, with $V$ and $A$ denoting the anomalous vector and axial gauge fields, and $W$ representing an extra non-anomalous gauge field. The remaining parameter $\kappa$ denotes the Chern-Simons coupling associated with the anomalous sectors. The Levi-Civita symbol is defined by $\epsilon(rtxyz) = +1$ whereas the Levi-Civita tensor is defined by $\epsilon^{MNPQR}=\epsilon(MNPQR)/\sqrt{-g}$. The Gibbons-Hawking-York boundary term is given by
		\begin{equation}
			\mathcal S_{\rm GH} = \dfrac{1}{8 \pi G}\int_\partial d^4x    \sqrt{h} K \,,
		\end{equation}
		and it is required to have a well-defined variational problem. The value of the coupling $\kappa$ is given in terms of the anomalous coefficient of the field theory as~\cite{Landsteiner:2011iq}
		\begin{equation}
			\label{eq:kappa-lambda_value}
			\frac{1}{16\pi G}\kappa = -\frac{1}{16\pi^2} \,.
		\end{equation}
		We will omit in this work a mixed gauge-gravitational Chern-Simons term that is sometimes taken into account in similar studies, see e.g. Refs.~\cite{Landsteiner:2011iq,Landsteiner:2012kd,Tamang:2025ekj}, as we are going to work in the probe limit so that its contribution would be vanishing. Notice that the gauge field $W$ does not appear at all in the Chern-Simons term, so that this symmetry is non-anomalous.  Next, we will further extend our model by introducing a symmetry breaking term, enabling us to investigate how quantum anomalies affect the non-anomalous sector of the theory rather than just the anomalous one.

		\subsection{Holographic model with symmetry breaking}
		\label{sec:sym_break}

		One of the main objectives of this work is to examine the possibility that the constitutive relation for the current $\langle J_w^\mu \rangle$ receives anomalous contributions despite the symmetry $w$ being non-anomalous. To explore this possibility, we extend our model of Eq.~(\ref{act}) by introducing symmetry breaking via a scalar field, $\mathcal S_\phi$ \cite{Megias:2016ery,Tamang:2025ekj,Tamang:2026epo}, so that the total action is
		\begin{equation}
			\mathcal S_{\textrm{tot}} = \mathcal S + \mathcal S_\phi  \,,
		\end{equation}
		where $\mathcal S$ is the bulk action comprising the usual Maxwell kinetic terms and the Chern-Simons contribution, while the additional scalar contribution is
		\begin{equation}
			\mathcal S_\phi = \frac{1}{16\pi G} \int d^5x \sqrt{-g} \bigg(  - |D_M \phi|^2 - m^2\phi^2  \bigg) \,, \qquad  D_M \phi = \left[ \partial_M -i(A_M - W_M) \right] \phi  \,. \label{eq:Sphi}
		\end{equation} 
		Here $\phi$ is a scalar field with a tachyonic bulk mass $m^2 = \Delta (\Delta -4) / \ell^2$, and $0 \le \Delta \le 4$. $\mathcal S_\phi$ produces a breaking of $a$ and $w$ symmetries via the scalar field~$\phi$. From the AdS/CFT dictionary, the model is the holographic dual of a Conformal Field Theory (CFT) with a deformation
		\begin{equation}
			{\cal L} = {\cal L}^{\rm CFT} +  \xi \, {\cal O} \,,
		\end{equation}
		where ${\cal O}$ is an operator dual of the scalar field with $\dim {\cal O} = \Delta$, and $\xi$ is the source of the operator with $\dim \xi = 4-\Delta$. The near boundary expansion of~$\phi$ reads
		\begin{equation}
			\phi(r) = \phi_{4-\Delta} \left( \frac{r}{\ell} \right) ^{\Delta-4} + \phi_{\Delta} \left( \frac{r}{\ell} \right)^{-\Delta} + \cdots \,, \qquad r \to \infty \,, \label{eq:phir}
		\end{equation}
		where, in standard quantization, $\phi_{4-\Delta}$ is interpreted as the source $\xi$, and $\phi_\Delta$ as the condensate $\langle {\cal O}\rangle$. In the following we will choose $\Delta = 3$, so that the bulk mass is $m^2 = -3/\ell^2$.

		\subsection{Background }
		\label{sec:back_sol}

		We will work in the probe limit, neglecting metric fluctuations hereafter. This approximation is justified as long as we focus on the study of chiral magnetic effects. In contrast, chiral vortical and energy transport effects would demand the inclusion of backreaction (see e.g. Ref.~\cite{Tamang:2025ekj}).
		
		\subsubsection{Ansatz and numerical solution}
                \label{sec:ansatz}
		
		We will consider the following ansatz for the metric
		\begin{equation}
			ds^2 = - \frac{r^2}{\ell^2}  f(r)dt^2 + \frac{\ell^2}{r^2} \frac{dr^2}{f(r)} + \frac{r^2}{\ell^2} \left( dx^2 + dy^2 + dz^2\right)   \,, \qquad f(r) = 1 - \frac{r_h^4}{r^4} \,,
		\end{equation}
		with Hawking temperature given by $T = \frac{r_h^2}{4\pi \ell^2} |f^\prime(r_h)| = \frac{r_h}{\pi \ell^2}$ where $r_h$ is the position of the horizon. For the background gauge and scalar fields we use
		\begin{equation}
			V = V_t(r)dt \,, \qquad A = A_t(r)dt \,, \qquad W = W_t(r)dt  \,, \qquad \phi = \phi(r) \,.
		\end{equation}
		The chemical potentials are computed as $\mu_s \equiv S_t(r\to\infty) - S_t(r_h)$ with $S = V, A, W$. By taking into account that the gauge fields must vanish on the horizon $S_t(r_h)=0$ $(S = V, A, W)$ as a consequence of regularity of the solution, these fields behave near the boundary as
		\begin{equation}\lim_{r\to\infty} V_t(r) = \mu_v  \,, \qquad \lim_{r\to\infty} A_t(r) = \mu_a  \,, \qquad  \lim_{r\to\infty} W_t(r) = \mu_w  \,. \label{eq:nearboundary}
		\end{equation}
		It is convenient to define the following fields $A_\pm \equiv A \pm W$, so that the covariant derivative writes~$D_M\phi = \left[ \partial_M -i A_{M-} \right] \phi$. Consequently, the chemical potentials take the form $\mu_\pm \equiv \mu_a  \pm \mu_w$. In the rest of the manuscript, we will set $\ell = 1$.

In the following we will consider the change of variable $u = r_h^2 / r^2$. In the $u$ variable the blackening factor is given by $f(u) = 1 - u^2$, and the horizon is located at $u_h = 1$. The equations of motion of the background for the fields $(V, A_+, A_-,\phi)$  are given by~\cite{Megias:2016ery,Tamang:2025ekj,Tamang:2026epo}
		\begin{eqnarray}
			0 &=& V_t^{\prime\prime} \,,  \label{eq:Vt}\\
			0 &=& A_{t+}^{\prime\prime} \,,  \label{eq:Atp}\\
			0 &=& A_{t-}^{\prime\prime} - \frac{ \phi^2}{u^2 f} A_{t-} \,,  \label{eq:Atm}\\
			0 &=& \phi^{\prime\prime} - \left( \frac{1}{u} - \frac{f^\prime}{f} \right) \phi^\prime + \frac{1}{4 u^2 f^2} \left( 3f +  u  \frac{A_{t-}^2}{r_h^2} \right) \phi \,, \label{eq:phi}
		\end{eqnarray}
		where primes denote derivatives with respect to $u$. These equations show that the background decouples into three distinct sectors: the $V$ sector, the $A_+$ sector, and the $(A_-,\phi)$ sector. The solution of $V_t$ and $A_{t+}$ can be obtained analytically, and the result is
		\begin{equation}
			V_t(u) = \mu_v \left( 1 - u  \right) \,, \qquad A_{t+}(u) = \mu_+ \left( 1 - u  \right)  \,.
		\end{equation}
		The solutions of $A_{t-}(u)$ and $\phi(u)$ can be obtained numerically by solving the coupled system of differential equations, Eqs.~(\ref{eq:Atm})-(\ref{eq:phi}), and using the boundary condition $A_{t-}(u_h) = 0$. We will solve this system of equations by using the {\it shooting method}: we will shoot from a point near the horizon $(u \simeq u_h)$, and find the numerical solution for the all regime up to a point near the boundary $(u \simeq 0)$. The near horizon expansions of the fields $A_{t-}$ and $\phi$ turn out to be
\begin{eqnarray}
A_{t-}(u) &=&  \zeta \times \left[  (u - 1) - \frac{\phi_h^2}{4} (u - 1)^2  + \mathcal O \left( (u - 1)^3 \right)  \right]  \,,  \label{eq:Atm_uv}\\
\phi(u) &=&  \phi_h \times \left[ 1 + \frac{3}{8} (u - 1) - \frac{(39  + 4 (\zeta/r_h)^2)}{256} (u - 1)^2 + \mathcal O\left( (u-1)^3\right) \right]  \,,  \label{eq:phi_uv}  
\end{eqnarray}
while the near boundary expansion of these fields is
		\begin{eqnarray}
			A_{t-}(u) &=& \mu_- - \frac{\rho_-}{r_h^2} u  + \cdots   \,,  \label{eq:Atm_uv_u}\\
			\phi(u) &=& \frac{O_1}{r_h} \, u^{1/2} + \frac{O_2}{r_h^3} \, u^{3/2} + \cdots \label{eq:phi_uv_u}  \,.
		\end{eqnarray}
		In these expressions $\phi_h$ is the value of the scalar field at the horizon, while $\mu_-$ and $\rho_-$ are the chemical potential and charge density, respectively. The chosen value $m^2 = -3$ allows to define two different theories: {\it i)} the {\it standard quantization} corresponds to a theory in which $O_1$ is interpreted as a coupling and $O_2$ as the expectation value of an operator with mass dimension~$3$, and {\it ii)} the {\it alternative quantization} in which $O_2$ is interpreted as the coupling and $O_1$ as the expectation value of an operator with mass dimension~$1$. In the following we will treat both theories, and we will refer to them as {\it $O_2$ theory} and {\it $O_1$ theory}, respectively. SSB in the $O_1$ theory [$O_2$ theory] is then produced by a non-vanishing value of the condensate $O_1 \ne 0$ $[O_2 \ne 0]$, when the source is vanishing $O_2 = 0$ $[O_1 = 0]$.

\begin{figure}[htbp]
   \centering
   \includegraphics[width=.46\textwidth]{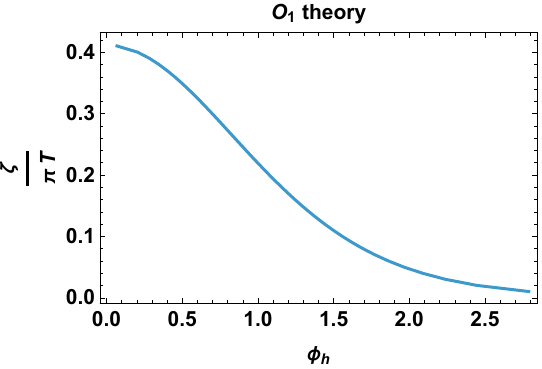} \hspace{1cm} \includegraphics[width=.46\textwidth]{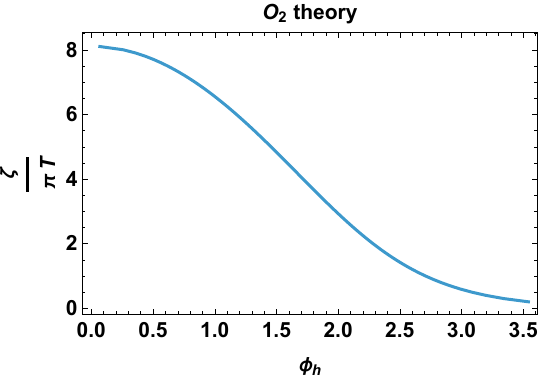}
   \caption{Plots of the relation between the near horizon parameters, $\phi_h$ and $\zeta/(\pi T)$, for the $O_1$ theory (left panel) and the $O_2$ theory (right panel) with SSB.}
\label{fig:zeta_phih}
\end{figure}

We have solved Eqs.~(\ref{eq:Atm})-(\ref{eq:phi}) numerically by shooting from a value $u_{\textrm{IR}} = 1 - \epsilon$ with $\epsilon \ll 1$, and using the expressions of Eqs.~(\ref{eq:Atm_uv})-(\ref{eq:phi_uv}) for the values of the fields at the shooting point. After fixing to some particular values the parameters of the solution near the horizon $(\zeta, \phi_h)$, we can extract from the near boundary expansion the corresponding values of the physical parameters $(\mu_-, \rho_-, O_1, O_2)$. This procedure leads to a relation between the near boundary and the near horizon quantities, i.e.
		\begin{equation}
			X = X(\zeta,\phi_h) \,, \qquad \textrm{for} \qquad X = (\mu_-, \rho_-, O_1, O_2) \,.
		\end{equation}
The next step is to fix either $O_1$ or $O_2$ to be vanishing, i.e.
		\begin{equation}
			O_i(\zeta,\phi_h) = 0\,, \qquad (i = 1, 2) \,.
		\end{equation}
		This condition produces a dependence between the parameters near the horizon
		\begin{equation}
			\zeta = \zeta(\phi_h)   \,,
		\end{equation}
                and consequently, one ends up with a single independent parameter in the solution near the horizon, which we choose to be~$\phi_h$. We display in fig.~\ref{fig:zeta_phih} the dependence $\zeta(\phi_h)$ for both theories: $O_1$ and $O_2$ theories; considering SSB.

We will present now the results of the numerical solutions of the background in the two theories with SSB.

\subsubsection{Results in the $O_1$ theory}
\label{sec:theoryO1}

\begin{figure}[htbp]
			\centering
			\includegraphics[width=.46\textwidth]{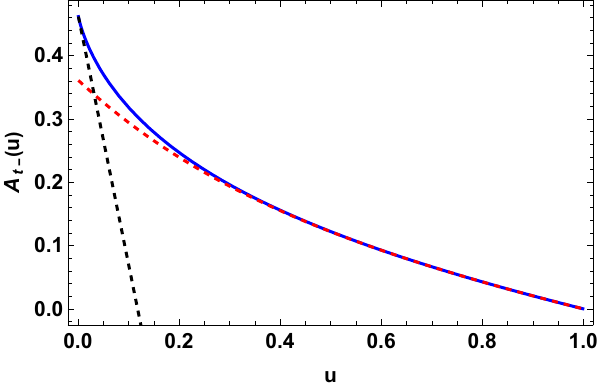}
			\qquad
			\includegraphics[width=.46\textwidth]{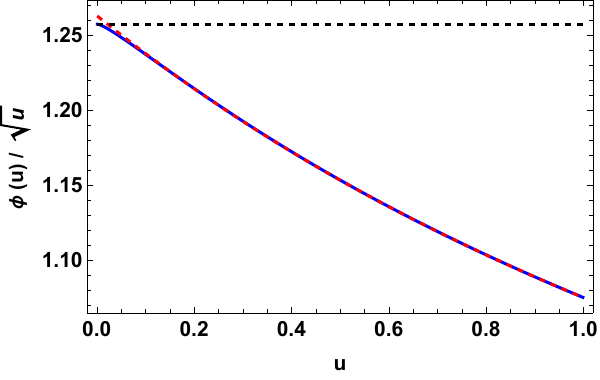}
			\caption{{\bf Results for the $O_1$ theory.} Plots of the solution of the background for $A_{t-}(u)$ (left panel) and for $\phi(u)$ (right panel) as a function of $u$. We display:  {\it (i)} the numerical solutions (solid blue lines), {\it (ii)} the analytical approximations near the horizon given by Eqs.~(\ref{eq:Atm_uv})-(\ref{eq:phi_uv}) (dashed red line), and {\it (iii)} the analytical approximations near the boundary given by Eqs.~(\ref{eq:Atm_uv_u})-(\ref{eq:phi_uv_u}) (dashed black lines). We have set $\zeta/(\pi T) = 0.2$, $\phi_h = 1.074$, $\bar\mu_- = 0.465$, $\rho_- / (\pi T)^3 = 3.942$, $\langle O_1\rangle / (\pi T) = 1.256$ and $\langle O_2\rangle/(\pi T)^3 = 0$.}
			\label{fig:back_sol1}
		\end{figure}

We consider in this subsection the $O_1$ theory with SSB, so that $ O_2 = 0$. We display in fig.~\ref{fig:back_sol1} the numerical solution of $A_{t-}(u)$ and $\phi(u)$ for fixed values of the parameters $(\zeta/(\pi T),\phi_h)$ computed in this theory. We can see that the numerical solutions correctly reproduce the two asymptotic regions: near horizon and near boundary regimes.

\begin{figure}[htbp]
		  \centering
                  			\includegraphics[width=.3\textwidth]{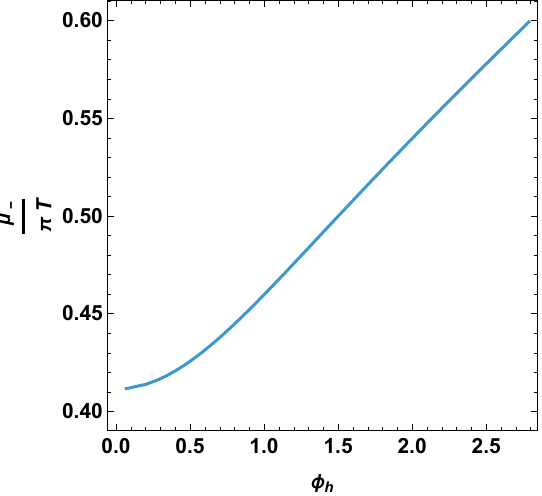}  \hspace{0.5cm} \includegraphics[width=.3\textwidth]{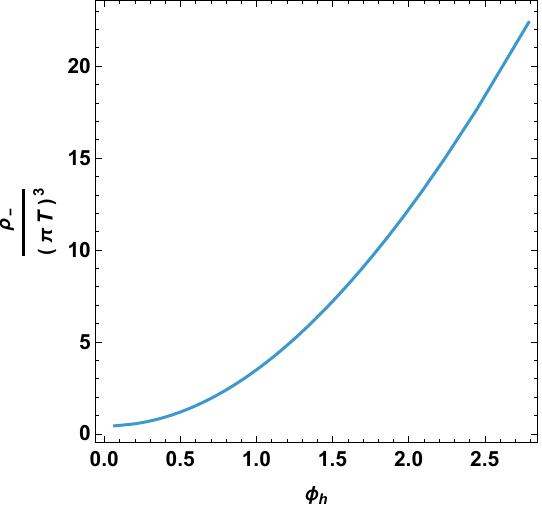} \hspace{0.5cm}
                  			\includegraphics[width=.3\textwidth]{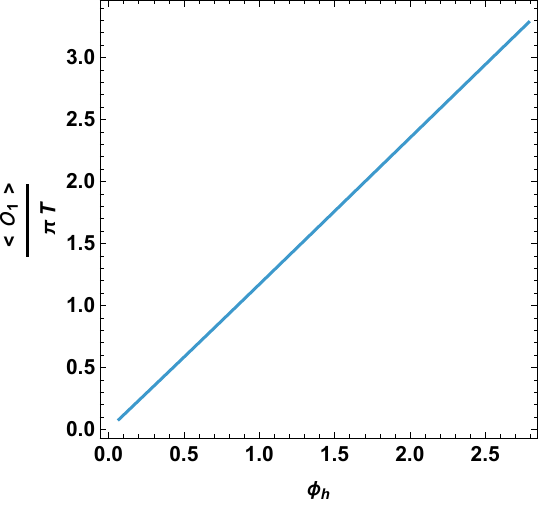} \hspace{0.5cm}  \\
\includegraphics[width=.3\textwidth]{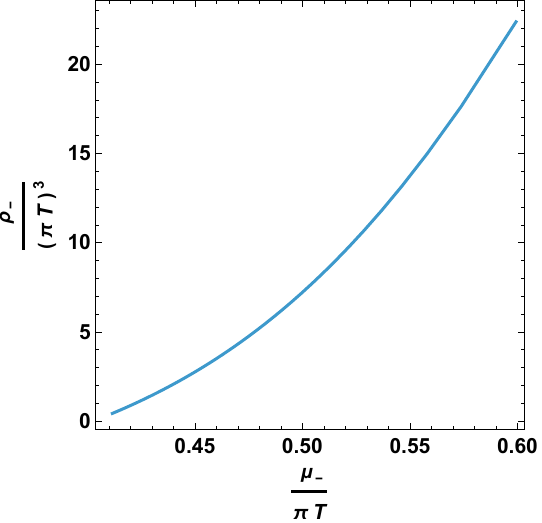} \hspace{0.5cm} \includegraphics[width=.3\textwidth]{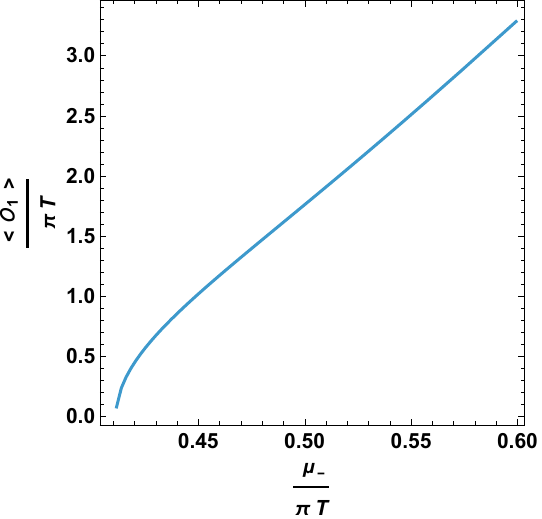}  \hspace{0.5cm} \includegraphics[width=.3\textwidth]{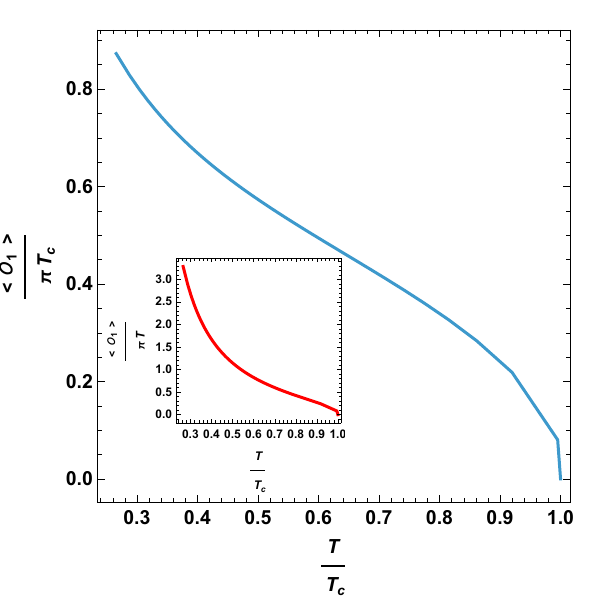} 
\caption{{\bf Results for the $O_1$ theory.} Upper panels: plot of $\mu_-/(\pi T)$ (left panel), $\rho_- / (\pi T)^3$ (middle panel) and $\langle O_1\rangle/(\pi T)$ (right panel) as a function of $\phi_h$. Lower panels: plot of~$\rho_-/ (\pi T)^3$ (left panel) and $\langle O_1 \rangle / (\pi T)$ (middle panel) as a function of $\mu_-/(\pi T)$, and $\langle O_1 \rangle/(\pi T_c)$ (right panel -- main plot) and $\langle O_1 \rangle/(\pi T)$ (right panel -- inserted figure) as a function of $T / T_c$.}
\label{fig:O1_theory}
\end{figure}

We show in fig.~\ref{fig:O1_theory} the behavior of the physical quantities $(\mu_-, \rho_-, O_1)$.
                We show in the upper panels the dependence of these quantities as a function of $\phi_h$. This dependence can be used to obtain $\rho_-(\mu_-)$, as it is displayed in the bottom-left panel of fig.~\ref{fig:O1_theory}. In the numerical analysis, we set $r_h \equiv \pi T =  1$. If we define the dimensionless quantities
                \begin{equation}
 \bar \mu_s \equiv \frac{\mu_s}{\pi T}\,,  \qquad \left( s \in \left\{v, + , - \right\} \right) \,,
                \end{equation}
then different values of $\bar\mu_s$ can be interpreted as either fixing the temperature $T$ while varying the chemical potential $\mu_s$, or vice versa.~\footnote{For clarity, the notation for dimensionless quantities, such as $\bar\mu_s \equiv \mu_s/(\pi T)$ and $\rho_s / (\pi T)^3$, will be maintained throughout the manuscript.} On the other hand, in $D = 5$ space-time dimensions, the critical temperature is proportional to $T_c \propto \rho_-^{1/3}$~\cite{Horowitz:2008bn}. Then, the dimensionless ratio $T / T_c$ can be written in the form
		\begin{equation}
			\frac{T}{T_c} = \alpha \frac{r_h}{\rho_-^{1/3}} \,, \label{eq:TTc} 
		\end{equation}
		where $\alpha$ is a dimensionless constant, and $r_h$ can be set to $1$ as mentioned above. From the numerical analysis, we find $\alpha \simeq 0.74889$. The expression of Eq.~(\ref{eq:TTc}) has been used to generate the plots in the lower panels of fig.~\ref{fig:O1_theory}. Finally, by fitting the curve of the condensate $\langle O_1 \rangle$ as a function of $T / T_c$, we observe that for small values of the condensate, its behavior follows a square-root dependence which is characteristic of a second-order phase transition. More specifically, as $ T \to T_c $ we find
		\begin{equation}
			\langle O_1 \rangle  \simeq 2.40426 \, T_c \left(  1 - \frac{T}{T_c} \right)^{1/2}  \,, \qquad
		\end{equation}
		where the critical temperature in this theory is $T_c = \rho_-^{1/3} / (\pi \alpha) \simeq 0.42504 \, \rho_-^{1/3}$.

\subsubsection{Results in the $O_2$ theory}
\label{sec:theoryO2}

\begin{figure}[htbp]
			\centering
			\includegraphics[width=.46\textwidth]{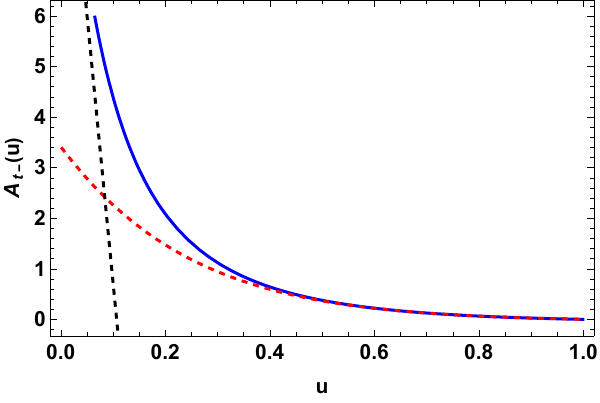}
			\qquad
			\includegraphics[width=.46\textwidth]{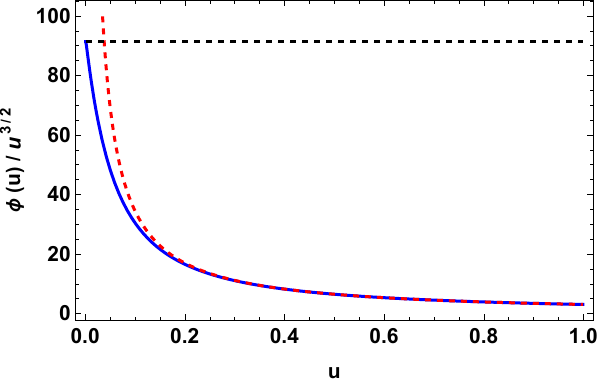}
			\caption{{\bf Results for the $O_2$ theory.} Plots of the solution of the background for $A_{t-}(u)$ (left panel) and for $\phi(u)$ (right panel) as a function of $u$. We have set $\zeta/(\pi T) = 0.2$, $\phi_h = 3.004$, $\bar\mu_- = 11.334$, $\rho_- / (\pi T)^3 = 105.951$, $\langle O_1\rangle / (\pi T) = 0$ and $\langle O_2\rangle/(\pi T)^3 = 91.368$. See fig.~\ref{fig:back_sol1} for further details.}
\label{fig:back_sol2}
		\end{figure}

\begin{figure}[htbp]
			\centering
\includegraphics[width=.3\textwidth]{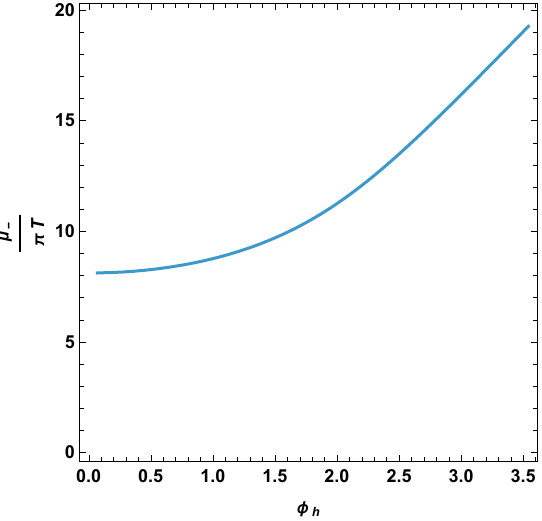}  \hspace{0.5cm} \includegraphics[width=.3\textwidth]{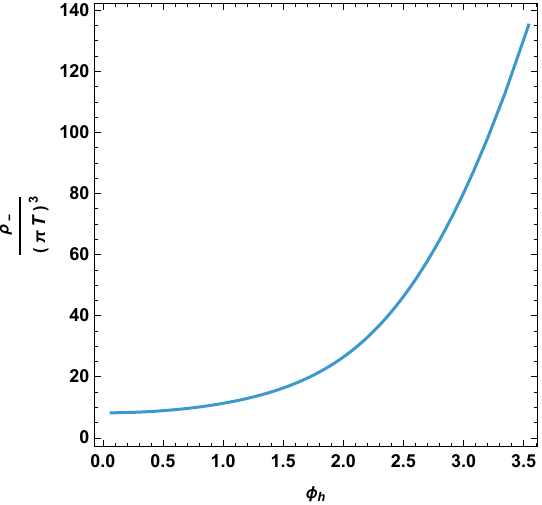}
\includegraphics[width=.3\textwidth]{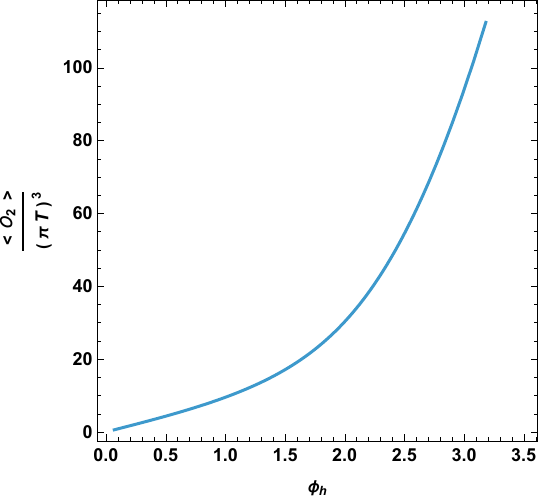}
 \\
\includegraphics[width=.3\textwidth]{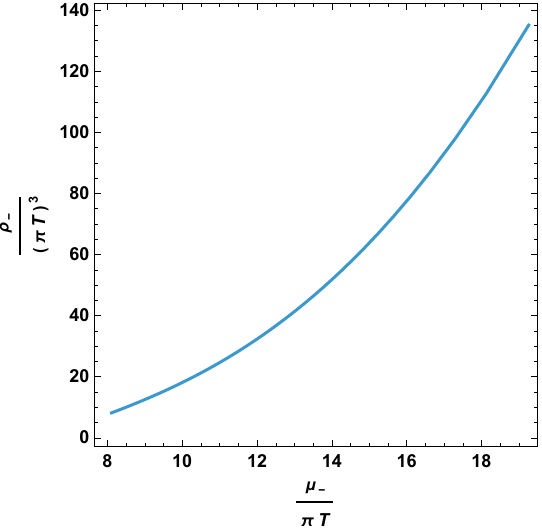} \hspace{0.5cm} \includegraphics[width=.3\textwidth]{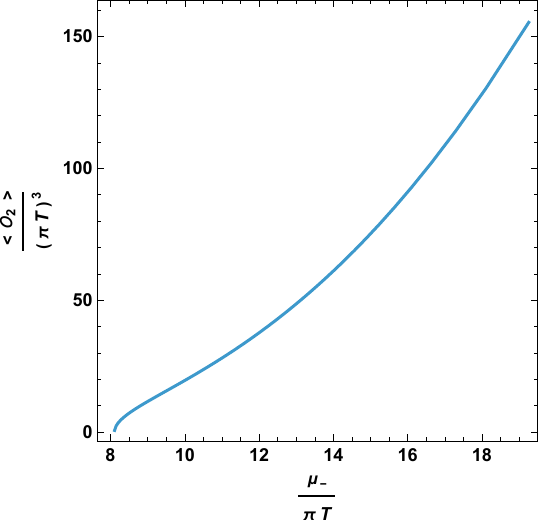}  \hspace{0.5cm} \includegraphics[width=.3\textwidth]{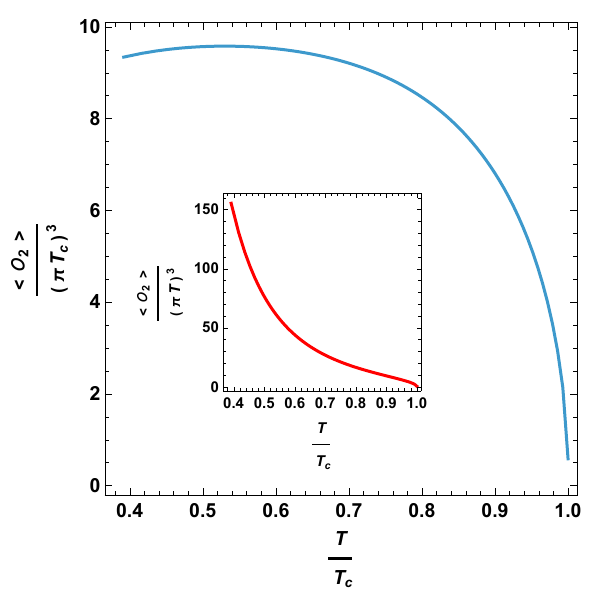}  \\

\caption{{\bf Results for the $O_2$ theory.} Upper panels: plot of $\mu_-/(\pi T)$ (left panel), $\rho_- / (\pi T)^3$ (middle panel) and $\langle O_2\rangle/(\pi T)^3$ (right panel) as a function of $\phi_h$. Lower panels: plot of~$\rho_-/ (\pi T)^3$ (left panel) and $\langle O_2 \rangle / (\pi T)^3$ (middle panel) as a function of $\mu_-/(\pi T)$, and $\langle O_2 \rangle/(\pi T_c)^3$ (right panel -- main plot) and $\langle O_2 \rangle/(\pi T)^3$ (right panel -- inserted figure) as a function of $T / T_c$. }
			\label{fig:O2_theory}
		\end{figure}

We show in this subsection the corresponding results for the $O_2$ theory with SSB, so that $O_1 = 0$. We display in fig.~\ref{fig:back_sol2} the numerical solution of $A_{t-}(u)$ and $\phi(u)$ for fixed values of the parameters $(\zeta/(\pi T),\phi_h)$, and in fig.~\ref{fig:O2_theory} the behavior of the physical quantities $(\mu_-, \rho_-, O_2)$. From the numerical analysis, the value of $\alpha$ in this theory is $\alpha \simeq 2.01011$, and the behavior of the condensate as $ T \to T_c$ is
		\begin{equation}
			\langle O_2 \rangle\simeq 733.55 \, T_c^3 \left(  1 - \frac{T}{T_c} \right)^{1/2}  \,,  \qquad 
		\end{equation}
		where the critical temperature in this case is  $T_c \simeq 0.158354 \, \rho_-^{1/3}$. The characteristic behavior of a second-order phase transition is also present in the $O_2$ theory.

            
                
		\section{Fluctuations}
		\label{sec:flut}
                
		We will now consider perturbations of the gauge fields around the numerical background solutions obtained in section~\ref{sec:back_sol}. To study the effect of quantum anomalies, we decompose the gauge fields into their background and linear perturbation parts,
		\begin{align}
			\mathcal{S}_\mu = S_\mu + \epsilon\, s_\mu \,,
		\end{align}
		where $S_\mu$ denotes the background solution for any of the fields $V$, $A_\pm$ (or, equivalently, $V$, $A$ and $W$), while $s_\mu$ denotes the corresponding fluctuations $v$, $a_\pm\equiv a\pm w$ (or, equivalently, $v$, $a$ and $w$). We then perform a Fourier-mode decomposition~\cite{Amado:2011zx},
		\begin{eqnarray}
			s_{\mu}(u,x^\nu)
			&=&
			\int \frac{d^4p}{(2\pi)^4}\,
			s_{\mu}(u)\,
			e^{-i\omega t+i\vec{p}\cdot\vec{x}}\,.
		\end{eqnarray}
		Without loss of generality, we choose the spatial momentum to be directed along the $z$-axis, $\vec{p}=p\,\hat{z}$. Since the conductivities are obtained from static correlators, we set $\omega=0$ and solve the fluctuation equations to linear order in $p\equiv|\vec{p}|$. The resulting gauge-field fluctuations determine the retarded current-current correlators entering the Kubo formulae. The explicit equations of motion for these fluctuations are given in appendix~\ref{sec:appendix_EoM_fluctuations}.

		\section{Constitutive relations and Kubo formulae}
		\label{sec:kubo_formulae}
		
		The transport coefficients for the chiral magnetic and vortical effects have been studied in a wide variety of methods which includes Kubo formulae \cite{Amado:2011zx,Landsteiner:2012kd,Jimenez-Alba:2014pea,Chowdhury:2015pba}, fluid gravity correspondence \cite{Bhattacharyya:2007vjd,Erdmenger:2008rm,Banerjee:2008th,Megias:2013joa}, diagrammatic methods \cite{Manes:2012hf} and the  partition function formalism \cite{Banerjee:2012iz,Megias:2014mba,Jensen:2012jy,Jensen:2012jh}. In this work we use Kubo formalism to study the chiral magnetic effect of anomalous conductivities. As explained above, we will study also the possibility that, under certain circumstances, this effect might be present in non-anomalous currents as well.
		
		In the following we will use the notation $J^i_s \in \{ J^i_v, \, J^i_a, \, J^i_w \}$, where we include the vector and axial currents, and the current related to the $w$ symmetry, respectively. The constitutive relations for the charged currents write
		\begin{eqnarray}
			\vec{J}_v &=& \sigma^B_{vv} \, \vec{B}_v + \sigma^B_{va} \, \vec{B}_a + \sigma^B_{vw} \, \vec{B}_w  \,,  \label{eq:Jv}  \\
			\vec{J}_a &=& \sigma^B_{av} \, \vec{B}_v + \sigma^B_{aa} \, \vec{B}_a + \sigma^B_{aw} \, \vec{B}_w \,,   \label{eq:Ja}   \\
			\vec{J}_w &=& \sigma^B_{wv} \, \vec{B}_v + \sigma^B_{wa} \, \vec{B}_a + \sigma^B_{ww} \, \vec{B}_w  \,.  \label{eq:Jw} 
		\end{eqnarray}
		We have denoted by $\sigma^B_{ss ^\prime}$ the chiral magnetic conductivity which induces the  current $\vec{J}_s$ with $s \in \{ v, a, w \}$  as a response of the magnetic field $\vec{B}_{s^\prime}$ associated to symmetry $s^\prime \in \{ v, a, w\}$.
		
		The Kubo formulae for the chiral magnetic conductivity in the dual field theory of the holographic model involves the retarded correlators of two currents, as it was derived in~\cite{Kharzeev:2009pj}. These formulae write
		\begin{equation}
			\sigma^B_{s s^\prime} = \lim_{p_k \rightarrow 0}\dfrac{i}{2 p_k}\sum_{i,j} \epsilon_{ijk}\langle J_s^{i} J_{s^\prime}^{j}\rangle\vert_{\omega =0} \,, \qquad   \left( s, s^\prime \in \{ v, a, w \} \right) \,.
			\label{eqku2}
		\end{equation}
		After taking the hermitian conjugate of the correlators, and using the hermiticity of the currents, one has that $\langle J_s^i(p) J_{s^\prime}^j(-p) \rangle = \langle J_{s^\prime}^j(p) J_s^j(-p) \rangle$ in the zero frequency limit. Then it can be deduced the following relations between the chiral magnetic conductivities~\cite{Tamang:2025ekj}
\begin{equation}
			\sigma^B_{s s^\prime} = \sigma^B_{s^\prime s}  \,, \qquad (s, s^\prime \in \{ v, a, w \} ) \,. \label{eq:sigmaB_symmetric}
\end{equation}
This property implies that the $3 \times 3$ conductivity $\sigma^B$ matrix, $(\sigma^B)_{s s^\prime}  \equiv \sigma^B_{s s^\prime}$, is symmetric.

		\section{Results for the conductivities}
		\label{sec:result}
		
		Before presenting the numerical results for the conductivities, we briefly outline the numerical method employed to solve the system of equations for the fluctuations detailed in appendix~\ref{sec:appendix_EoM_fluctuations}.
		
		\subsection{Numerical method}
		\label{subsec:numerical_method}
		
		To solve the coupled differential equations for the fluctuations we have used the pseudo-spectral method~\cite{trefethen2000spectral, boyd2000spectral}. To get the solution within this method, we expand the fluctuations as a sum of Chebyshev polynomials in the $u$ direction given by 
		\begin{equation}
			(a_\pm)_i=\sum_{m=0}^{N-1} (a_\pm^{m})_i \, T_m (2u-1) \,,  \quad
			v_i=\sum_{m=0}^{N-1}v_i^m \, T_m (2u-1) \, \,,
			\label{cheb}
		\end{equation}
		where $(a_\pm^m)_i$ and $v_i^m$  are the coefficients of the polynomials in (\ref{cheb}) with ($i=x,y$). To implement this procedure numerically, the computational domain $u \in [ 0,1 ]$ is discretized using a Chebyshev-Gauss-Lobatto grid. The grid points are distributed according to
                \begin{equation}
			u_j=\frac{1}{2}\left( 1-\cos\left(\frac{\pi j}{N-1}\right) \right) \,, \quad\quad j=0,1,\dots,N-1 \,,
		\end{equation}
where $N$ denotes the number of grid points, which equals the number of coefficients in Eq.~(\ref{cheb}). Within the pseudo-spectral method, the equations of motion are rewritten by substituting the differential operators with differentiation matrices. For a function $f$ evaluated on the grid, the derivatives are approximated via matrix multiplication,
		\begin{equation}
			f^{(n)}(u_i) \approx \sum_{j=0}^{N-1}D_{i j}^{(n)}f(u_j) \,,
		\end{equation}
		where $D^{(n)}$ represents the $n$-th order differentiation matrix. The fluctuation fields $(a_\pm)_i$ and $v_i$  are expanded in Chebyshev polynomials as in Eq.~(\ref{cheb}), with the expansion coefficients mapped to the field values at the collocation points.  Substituting the derivative matrices into the equations of motion gives an algebraic system for these grid values.  We then evaluate this system at the interior grid points $i=1, \dots, N-2$, and impose boundary conditions at the endpoints. The resulting algebraic system is solved to obtain the fluctuation profiles.
		
		
		In the following subsections we will present our numerical results for the chiral magnetic conductivities in the $O_1$ and $O_2$ theories with SSB. For comparison, we will display also for some of the conductivities the results obtained in the large condensates limit by using the analytical formulas that are derived in appendix~\ref{sec:sigma_analytic}. Notice that we are working at zero frequency and in the limit of zero momentum: $\omega=0$ and $p\rightarrow 0$.

		\subsection{Results of the conductivities in the $O_1$ theory}
		\label{sec:result1}
                
		Figs.~\ref{fig:fluct1} and \ref{fig:fluct2} present, as solid lines, the numerical results for the chiral magnetic conductivities appearing in the constitutive relations~(\ref{eq:Jv})-(\ref{eq:Jw}) within the $O_1$ theory, plotted as a function of $\langle O_1 \rangle/(\pi T)$. In these plots we have fixed $\bar\mu_v=0.1$ and $\bar\mu_+=0.5 $, while varying~$\bar\mu_-$. Choosing to fix the values of $(\bar\mu_v, \bar\mu_+)$, rather than $(\bar\mu_v,\bar\mu_a)$ or $(\bar\mu_v,\bar\mu_w)$, is motivated by the distinct sectors in which the background decouples, cf. section~\ref{sec:ansatz}. 

                We can see in these figures that two of the conductivities inducing the axial current $\vec{J}_a$ ($\sigma^{B}_{av}$ and $\sigma^{B}_{aa}$) are monotonically suppressed as $\langle O_1 \rangle / (\pi T)$ increases, whereas the conductivities driving the non-anomalous current $\vec{J}_w$ ($\sigma^{B}_{wv}$, $\sigma^{B}_{wa}$ and $\sigma^{B}_{ww}$) are enhanced. This indicates a redistribution of the magnetic response from the axial to the non-anomalous sector, while the vector sector remains less affected. Furthermore, as a consequence of the symmetry property in Eq.~(\ref{eq:sigmaB_symmetric}), all the chiral magnetic conductivities governing the response to a magnetic field in the non-anomalous sector ($\sigma^{B}_{sw}$ with $s \in \{v, a, w \}$) increase with $\langle O_1 \rangle / (\pi T)$, whereas $\sigma^B_{va}$ decreases. Finally, the conductivity $\sigma^B_{vv}$ tends to be less sensitive to variations of the condensate, although its precise behavior depends on the specific value of the chemical potential $\bar\mu_+$, as can be deduced from table~\ref{tab:sigma_analytic}. The behaviors of all these conductivities are qualitatively similar to those obtained in the presence of explicit symmetry breaking, which was studied in~\cite{Tamang:2025ekj}.

                In these figures, the large condensate limits of $\sigma^B_{va} = \sigma^B_{av}$ and $\sigma^B_{vw} = \sigma^B_{wv}$, obtained from the analytical formulas in table~\ref{tab:sigma_analytic}, are displayed as dashed horizontal lines. For the other conductivities, these formulas fail to accurately reproduce the numerical results because they neglect the $O_1$-dependence of the chemical potential~$\mu_-$. To illustrate this point, the left panel of fig.~\ref{fig:fluct1} includes the large $O_1$ limit for $\sigma^B_{vv}$, which fails to reproduce the numerical value by a factor of 2. Consequently, we choose to omit the corresponding lines for these conductivities in the remaining figures. In section~{\ref{subsec:disc_vv_aa_ww}, we will characterize  the functional dependence of all conductivities on both the condensate and the chemical potentials; by incorporating this $O_1$-dependence into $\mu_-$, the proposed parameterization will successfully describe their overall behavior.

		\begin{figure}[htbp]
			\centering
			\includegraphics[width=.43\textwidth]{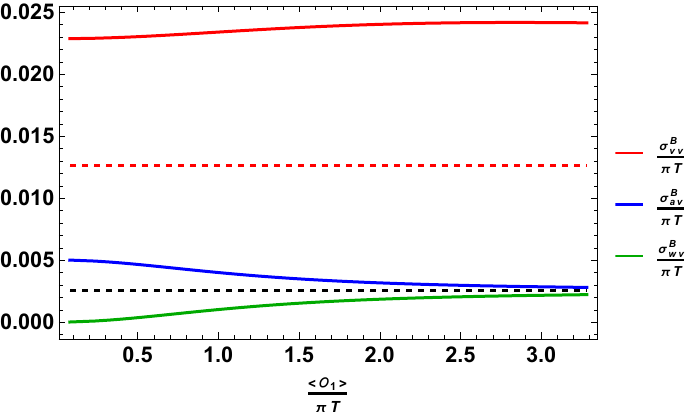}
			\includegraphics[width=.43\textwidth]{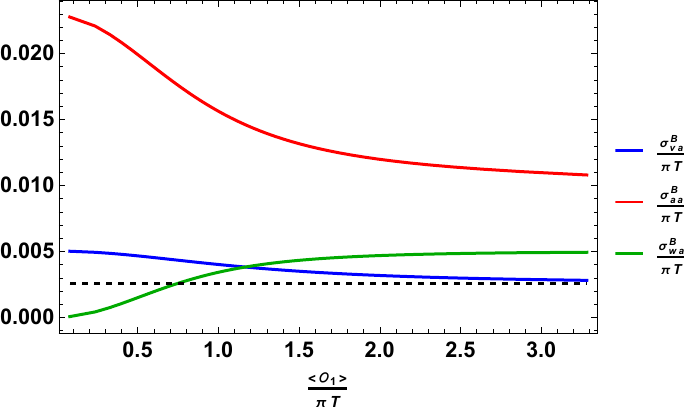}
			\caption{{\bf Results for the $O_1$ theory.} Plots of the conductivities $\sigma^{B}_{sv}$ (left panel) and $\sigma^{B}_{sa}$ (right panel), where $s \in \left\{ v,a,w \right\}$,  as a function of the condensate $\langle O_1 \rangle/ (\pi T)$. The dashed horizontal lines correspond to the analytic result in the $O_i \to \infty$ limit $(i = 1, 2)$ given in table~\ref{tab:sigma_analytic} for $\sigma^B_{va} = \sigma^B_{av}$ and $\sigma^B_{vw} = \sigma^B_{wv}$ (dashed black horizontal lines) and $\sigma^B_{vv}$ (dashed red horizontal line). We have set $\bar\mu_v=0.1$ and $\bar\mu_+=0.5 $, while varying~$\bar\mu_-$.}
               \label{fig:fluct1}
		\end{figure}
		
		\begin{figure}[htbp]
			\centering
			\includegraphics[width=.43\textwidth]{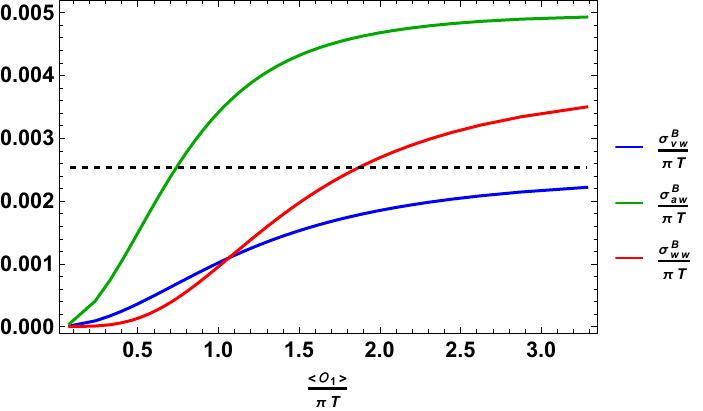}
			\caption{{\bf Results for the $O_1$ theory.} Plots of the conductivities $\sigma^{B}_{sw}$, where $s \in \left\{ v,a,w \right \}$, as a function of the condensate $\langle O_1 \rangle/ (\pi T)$. See fig.~\ref{fig:fluct1} for further details.}
                        \label{fig:fluct2}
		\end{figure}
		%


\subsection{Results of the conductivities in the $O_2$ theory}
\label{sec:result2}

Figs.~\ref{fig:fluct4}, \ref{fig:fluct5} and \ref{fig:fluct6} present the chiral magnetic conductivities within the $O_2$ theory as a function of $\langle O_2 \rangle / (\pi T)^3$. Similarly to the theory $O_1$, we have fixed $\bar\mu_v=0.1$ and $\bar\mu_+=0.5$ in these plots while varying~$\bar\mu_-$. The resulting conclusions align with those obtained for the $O_1$ theory, although minor differences appear in some diagonal terms of the conductivity matrix~$\sigma^B$. Specifically,~$\sigma^B_{aa}$ decreases for small values of the condensate but becomes monotonically increasing in the large condensate regime. As with the $O_1$ theory, we will characterize  the functional dependence of these conductivities in section~\ref{subsec:disc_vv_aa_ww}.
		
		\begin{figure}[htbp]
			\centering
			\includegraphics[width=.43\textwidth]{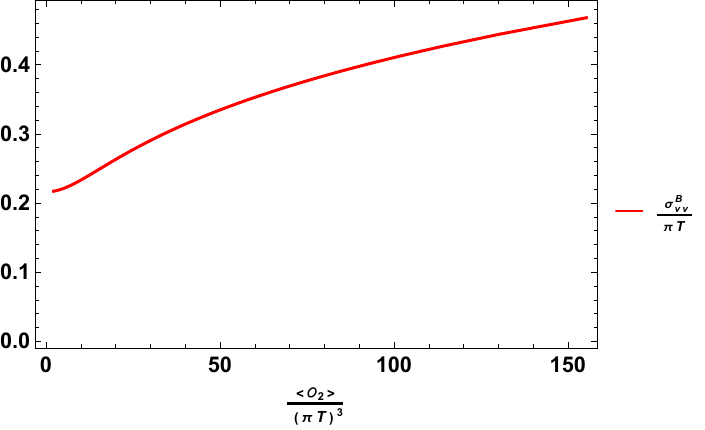}
			\includegraphics[width=.43\textwidth]{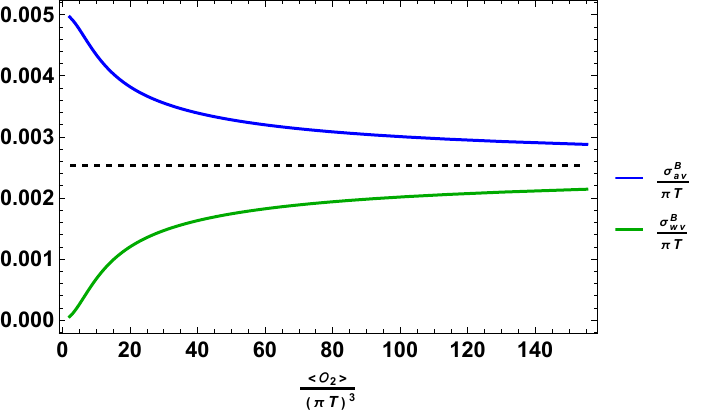}
			\caption{{\bf Results for the $O_2$ theory.} Plots of the conductivities $\sigma^{B}_{vv}$ (left panel), $\sigma^{B}_{av}$ and $\sigma^{B}_{wv}$ (right panel), as a function of the condensate $\langle O_2 \rangle/ (\pi T)^3$.  See fig.~\ref{fig:fluct1} for further details.}
 \label{fig:fluct4}
\end{figure}

		\begin{figure}[htbp]
			\centering
			\includegraphics[width=.43\textwidth]{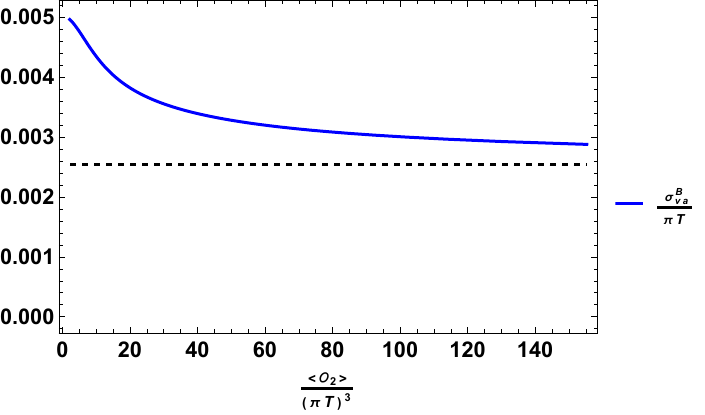}
			\includegraphics[width=.43\textwidth]{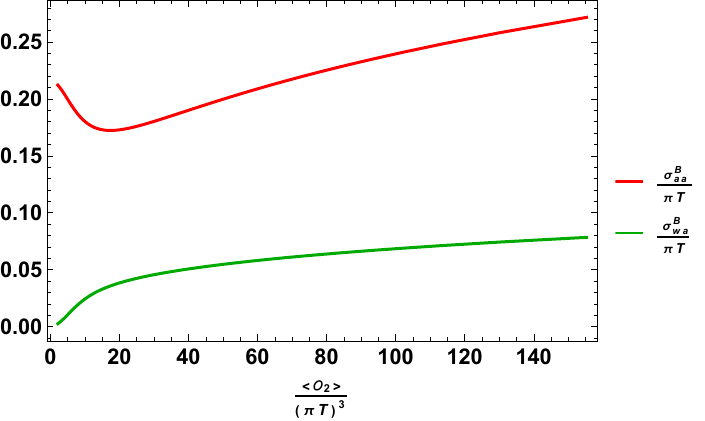}
			\caption{{\bf Results for the $O_2$ theory.} Plots of the conductivities $\sigma^{B}_{va}$ (left panel), $\sigma^{B}_{aa}$ and $\sigma^{B}_{wa}$ (right panel), as a function of the condensate $\langle O_2 \rangle/ (\pi T)^3$. See fig.~\ref{fig:fluct4} for further details.}
                \label{fig:fluct5}
		\end{figure}
		
\begin{figure}[htbp]
			\centering
			\includegraphics[width=.43\textwidth]{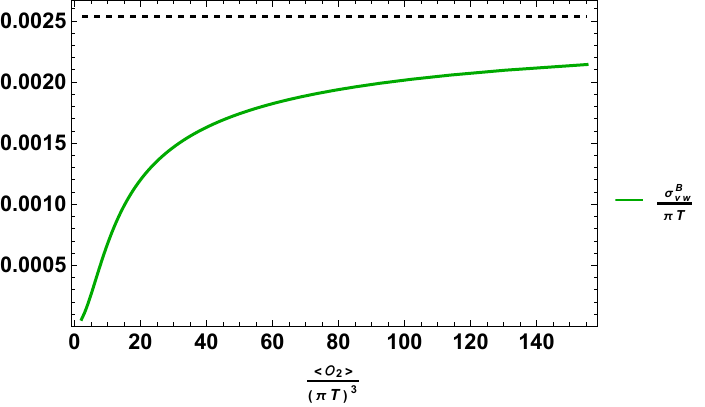}
			\includegraphics[width=.43\textwidth]{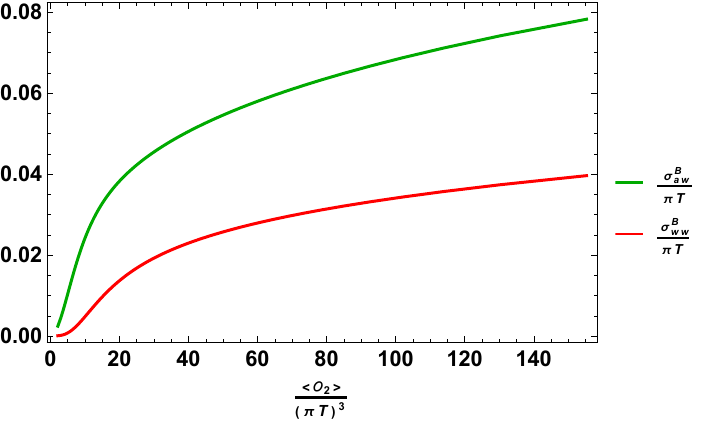}
			\caption{{\bf Results for the $O_2$ theory.} Plot of the conductivities $\sigma^{B}_{vw}$ (left panel), $\sigma^{B}_{aw}$ and $\sigma^{B}_{ww}$ (right panel), as a function of the condensate $\langle O_2 \rangle/ (\pi T)^3$. See fig.~\ref{fig:fluct4} for further details.}
\label{fig:fluct6}
\end{figure}

Finally, appendix~\ref{sec:appendix_sigma} also shows the behavior of all the chiral magnetic conductivities as a function of $T/T_c$ for both the $O_1$ and $O_2$ theories.

\subsection{Characterizing the dependence of the conductivities}
\label{subsec:disc_vv_aa_ww}

In this section, we characterize the dependence of the conductivities on the condensates and on the chemical potentials. We begin our analysis with the conductivities $\sigma^B_{va}$ and $\sigma^B_{vw}$, and then proceed to the remaining~ones.

\subsubsection{Characterizing $\sigma^B_{va}$ and $\sigma^B_{vw}$}

From the asymptotic behaviors of the conductivities shown in table~\ref{tab:sigma_analytic} of appendix~\ref{sec:sigma_analytic}, it can be seen that $\sigma^B_{va}(O_{i})$ decreases from $\mu_v/(2\pi^2)$ at $O_i = 0$, to $\mu_v / (4\pi^2)$ at $O_i \to \infty$. Consequently, for positive values of the chemical potential $\mu_v$, this conductivity is expected to be monotonically decreasing. A similar reasoning applies to $\sigma^B_{vw}(O_{i})$, suggesting that it becomes monotonically increasing as the condensates increase. These monotonic behaviors are confirmed by the numerical results presented in sections~\ref{sec:result1} and \ref{sec:result2}. Furthermore, inspired by these analytical formulas, we propose to describe the behavior of these conductivities through the following expressions
\begin{equation}
  \sigma^B_{va}(O_i) = \frac{\mu_v}{2 \pi^2} f^{(i)}_{va}(O_i)  \,, \qquad \sigma^B_{vw}(O_i) = \frac{\mu_v}{4 \pi^2} f^{(i)}_{vw}(O_i) \,,\label{eq:fivaw1}   
\end{equation}
where the functions $f_{va}^{(i)}$ and $f_{vw}^{(i)}$ satisfy the asymptotic limits
\begin{equation}
  f_{va}^{(i)}(0) = 1 \,, \qquad f_{va}^{(i)}(\infty) = \frac{1}{2} \,, \qquad f_{vw}^{(i)}(0) = 0 \,, \qquad f_{vw}^{(i)}(\infty) = 1 \,.  \label{eq:fva_vw_asymp}
\end{equation}
Fig.~\ref{fig:sigma_va_vw} displays the behavior of the normalized conductivities $(2\pi^2 / \mu_v)\times \sigma_{va}^{B}$ and $(4\pi^2/\mu_v) \times \sigma_{vw}^B$ as a function of the condensates $O_{1,2}$. This normalization isolates the condensate dependence of the functions $f^{(i)}_{va}(O_i)$ and $f^{(i)}_{vw}(O_i)$ for the $O_1$ and $O_2$ theories. As shown in the figure, the numerical behavior of these functions agree perfectly with the asymptotic limits of Eq.~(\ref{eq:fva_vw_asymp}). Furthermore, we have verified that these results are independent on the specific values of $\mu_v$ and $\mu_+$.

\begin{figure}[htbp]
	\centering
	\includegraphics[width=.44\textwidth]{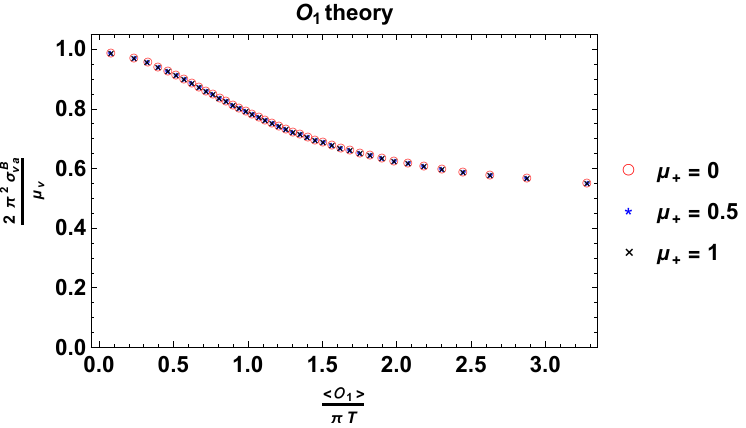}
	\includegraphics[width=.44\textwidth]{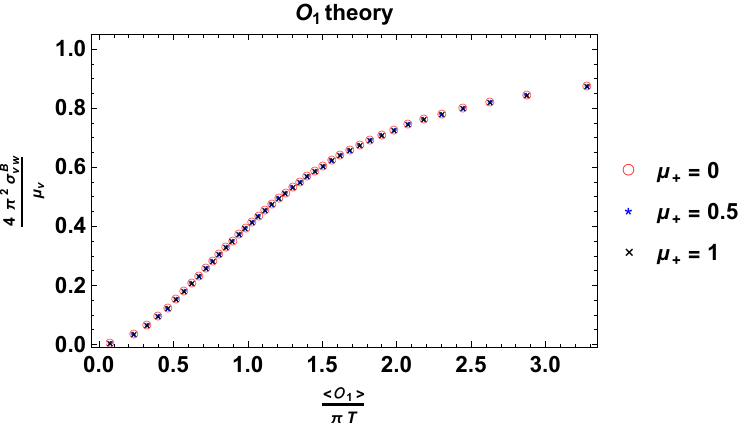}
	\includegraphics[width=.44\textwidth]{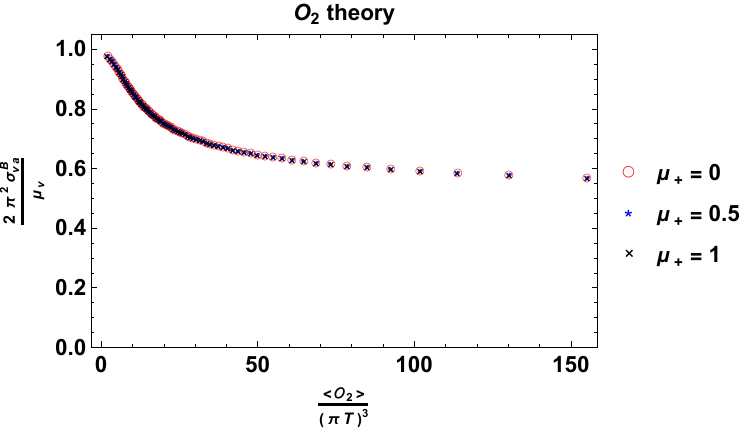}
	\includegraphics[width=.44\textwidth]{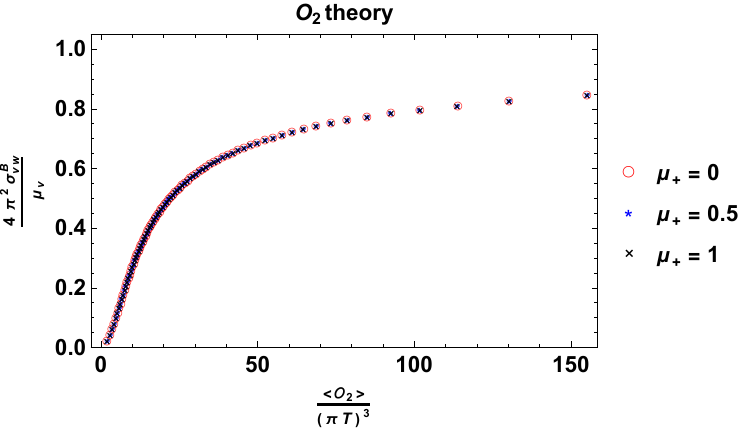}
	\caption{{\bf Results for the $O_1$ and $O_2$ theories.} Plots of the normalized conductivities $f_{va}^{(i)}(O_i) \equiv (2\pi^2  / \mu_v) \times \sigma^B_{va}(O_i)$ (left panels) and $f_{vw}^{(i)}(O_i) \equiv (4\pi^2/\mu_v) \times \sigma^B_{vw}(O_i)$ (right panels) ($i=1,2$), for $\mu_+ = 0$, $0.5$ and $1$, in the $O_1$ theory (upper panels) and the $O_2$ theory (lower panels).}
	\label{fig:sigma_va_vw}
\end{figure}

\subsubsection{Characterizing $\sigma^B_{vv}$, $\sigma^B_{aa}$, $\sigma^B_{ww}$ and $\sigma^B_{aw}$}

The situation is less straightforward for the conductivities $\sigma^B_{vv}$ and $\sigma^B_{aa}$, as they depend linearly on $\mu_-$ (at least for small condensate values, cf. table~\ref{tab:sigma_analytic}), and this chemical potential itself depends on the condensates~$O_{1,2}$. As shown below, a similar linear dependence on~$\mu_-$ emerges for $\sigma^B_{ww}$ and $\sigma^B_{aw}$ at intermediate values of the condensates. To examine these conductivities in details, we assume that their numerical behavior can be described by the following formulas~$(i=1,2)$
\begin{eqnarray}
\sigma^B_{vv}(O_i) &=& \frac{f^{(i)}_{v}(O_i) \mu_+ + g^{(i)}_{v}(O_i) \mu_-(O_i)}{4 \pi^2} \,,\label{eq:fivv1} \\
\sigma^B_{aa}(O_i) &=& \frac{f^{(i)}_{a}(O_i) \mu_+ + g^{(i)}_{a}(O_i)\mu_-(O_i)}{4 \pi^2} \,, \label{eq:fiaa1} \\
\sigma^B_{ww}(O_i)  &=& \frac{f^{(i)}_{w}(O_i) \mu_+ + g^{(i)}_{w}(O_i) \mu_-(O_i)}{16 \pi^2} \,,  \label{eq:fiww1} \\
\sigma^B_{aw}(O_i) &=& \frac{f^{(i)}_{aw}(O_i) \mu_+ + g^{(i)}_{aw}(O_i) \mu_-(O_i)}{16 \pi^2} \,. \label{eq:fiaw1}  
\end{eqnarray}
These expressions allow to interpolate between the two asymptotic limits, $O_i \to 0$ and $O_i \to \infty$. From a comparison of these formulas with table~\ref{tab:sigma_analytic}, one deduces that the functions appearing in Eqs.~(\ref{eq:fivv1})-(\ref{eq:fiaw1}) should have the asymptotic limits shown in table~\ref{tab:fg_limits}. For completeness, we also include the results from Eq.~(\ref{eq:fva_vw_asymp}) in this table. 
\begin{table}[htb]
\centering
\begin{tabular}{||c||c|c|c|c|c|c|c|c|c|c||}
\hline
& $f_v^{(i)}$  &  $g_v^{(i)}$   &  $f_a^{(i)}$  & $g_a^{(i)}$  & $f_w^{(i)}$  & $g_w^{(i)}$  & $f_{aw}^{(i)}$  & $g_{aw}^{(i)}$ & $f_{va}^{(i)}$ & $f_{vw}^{(i)}$  \\ 
\hline \hline
$O_i = 0$  &  1 &  1 & 1 & 1 & 0 & 0 & 0 & 0 & 1 & 0   \\
\hline
$O_i \to \infty$  &  1 &  0 & $\frac{1}{4}$ & 0 & 1 & 0 & 1 & 0 & $\frac{1}{2}$ & 1  \\
\hline
				\end{tabular} 
			\caption{\it Values of the functions $f_s^{(i)}(O_i)$ and $g_s^{(i)}(O_i)$, for $O_i = 0$ and $O_i \to \infty$  $(i=1,2)$.}
			\label{tab:fg_limits}
		\end{table}
The dependence of these functions with the condensates can be extracted from the numerical results of the conductivities by evaluating them  at different values of the chemical potential~$\mu_+$. In particular, from Eq.~(\ref{eq:fivv1}) one easily obtains $(i = 1,2)$
\begin{eqnarray}
f^{(i)}_{v}(O_i) &=& 4 \pi^2 \times \frac{  \sigma^B_{vv,1}(O_i) -  \sigma^B_{vv,2}(O_i) }{ \mu_{1,+} - \mu_{2,+} }  \,, \\
g^{(i)}_{v}(O_i) &=& \frac{4\pi^2}{\mu_-(O_i)}  \times \frac{ \mu_{1,+} \sigma^B_{vv,2}(O_1) - \mu_{2,+} \sigma^B_{vv,1}(O_i) }{ \mu_{1,+} - \mu_{2,+} } \,,
\end{eqnarray}
where $\mu_{1,+}$ and $\mu_{2,+}$ are two arbitrary values of the chemical potential $\mu_+$, while $\sigma^B_{vv,1}$ and $\sigma^B_{vv,2}$ are the corresponding conductivities evaluated at $\mu_+ = \mu_{1,+}$ and $\mu_{2,+}$, respectively. Similar expressions apply for the other conductivities.

The plots of the functions $f_s^{(i)}(O_i)$ and $g_s^{(i)}(O_i)$ are shown for $i=1$ in fig.~\ref{fig:fva1}, and for $i=2$ in fig.~\ref{fig:fva2}. Notably, these results become independent of $\bar\mu_v$ as well as the values assigned to $\bar\mu_{1,+}$ and $\bar\mu_{2,+}$. The numerical behavior $f^{(i)}_{v}(O_i) = 1$ $\forall O_i$ is in clear agreement with the asymptotic limits shown in table~\ref{tab:fg_limits}. For the remaining functions, the observed trends align with the asymptotic expectations; nevertheless, exploring larger values of $O_1/(\pi T)$ and $O_2 / (\pi T)^3$ would be necessary for a definitive validation. Finally, let us point out that the contributions $\propto g_v^{(i)}(O_i) \bar \mu_-(O_i)$ in Eq.~(\ref{eq:fivv1}) for $i = 1, 2$, increase for larger values of the condensates, even when $g_v^{(i)}(O_i)$  decrease; these terms are responsible for the growth of $\sigma^B_{vv} / (\pi T)$ observed in figs.~\ref{fig:fluct1} and \ref{fig:fluct4}. Similarly, the contribution $\propto g_a^{(2)}(O_2) \bar \mu_-(O_2)$ in Eq.~(\ref{eq:fiaa1}) also increases at large condensate values, thereby governing the non-monotonic behavior of $\sigma^B_{aa}/(\pi T)$ in fig.~\ref{fig:fluct5}.
\begin{figure}[htbp]
\centering
\includegraphics[width=.44\textwidth]{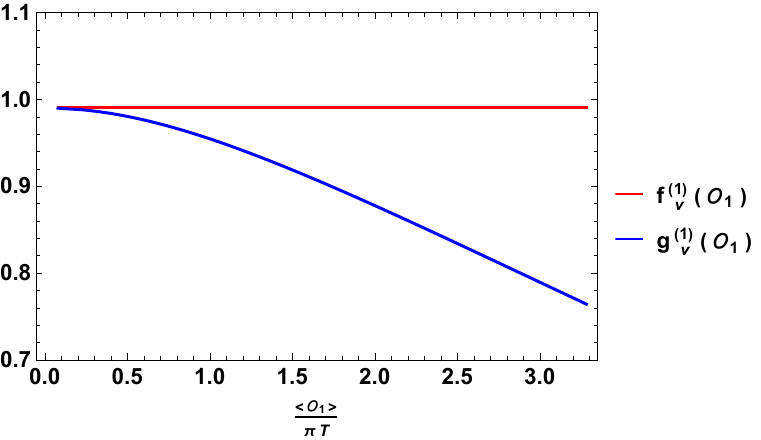}
\includegraphics[width=.44\textwidth]{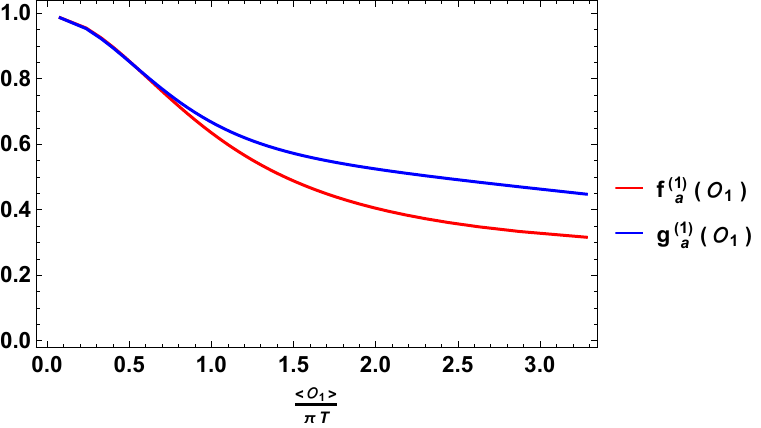}
\includegraphics[width=.44\textwidth]{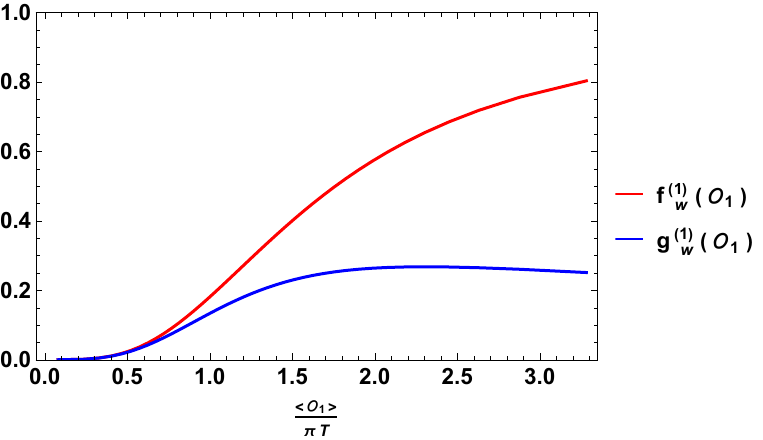}
\includegraphics[width=.44\textwidth]{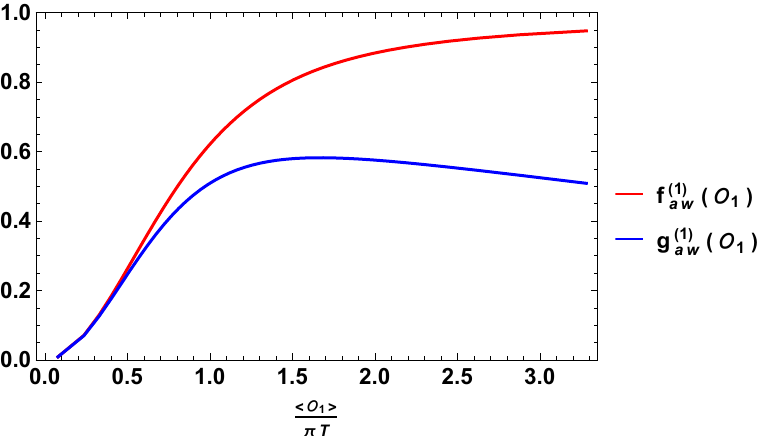}
\caption{{\bf Results for the $O_1$ theory.} Plots of the functions $f^{(1)}_{s}(O_1)$, $g^{(1)}_{s}(O_1)$, where $s \in \{ v, a, w\}$, $f^{(1)}_{a w}(O_1)$ and $g^{(1)}_{a w}(O_1)$, as a function of condensate $\langle O_1 \rangle/ (\pi T)$. Color code: Red -  $f^{(1)}_{v}(O_1)$, Blue - $g^{(1)}_{v}(O_1)$ (upper left panel);  Red -  $f^{(1)}_{a}(O_1)$, Blue - $g^{(1)}_{a}(O_1)$ (upper right panel): Red -  $f^{(1)}_{w}(O_1)$, Blue - $g^{(1)}_{w}(O_1)$ (lower left panel);  Red -  $f^{(1)}_{aw}(O_1)$, Blue - $g^{(1)}_{aw}(O_1)$ (lower right panel). We have used $\bar\mu_{1,+}=0.4$ and $\bar\mu_{2,+}=0.5$.
}
\label{fig:fva1}
\end{figure}
\begin{figure}[htbp]
\centering
\includegraphics[width=.44\textwidth]{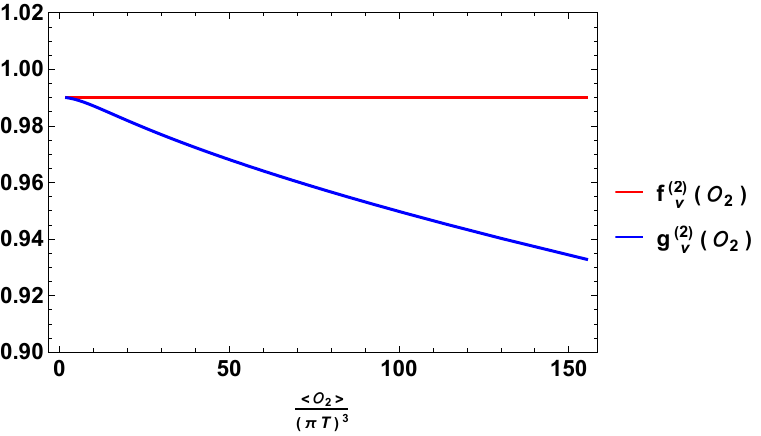}
\includegraphics[width=.44\textwidth]{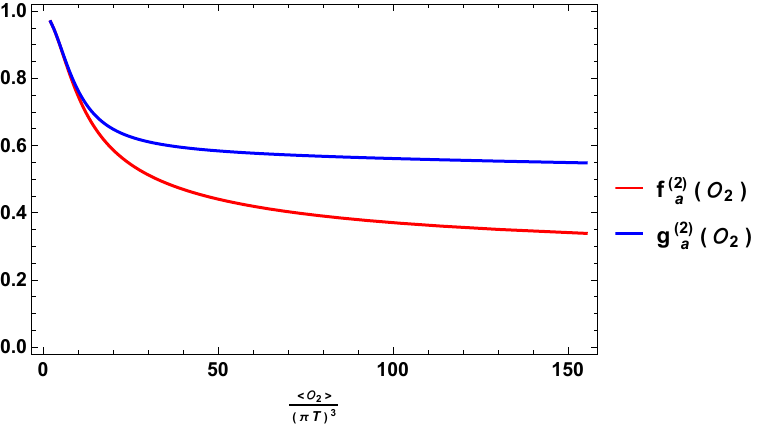}
\includegraphics[width=.44\textwidth]{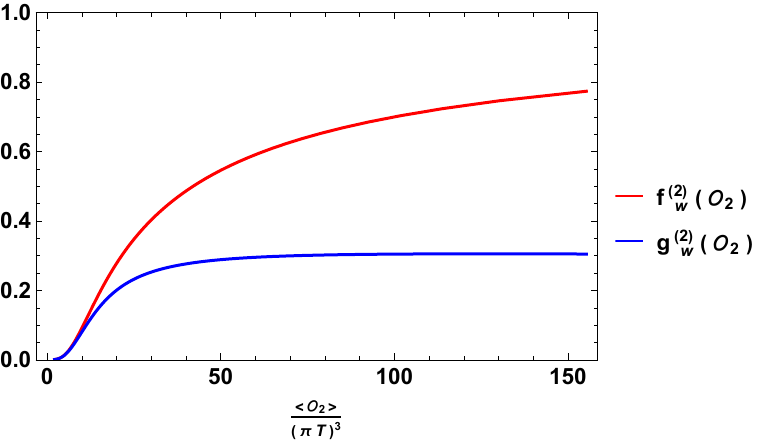}
\includegraphics[width=.44\textwidth]{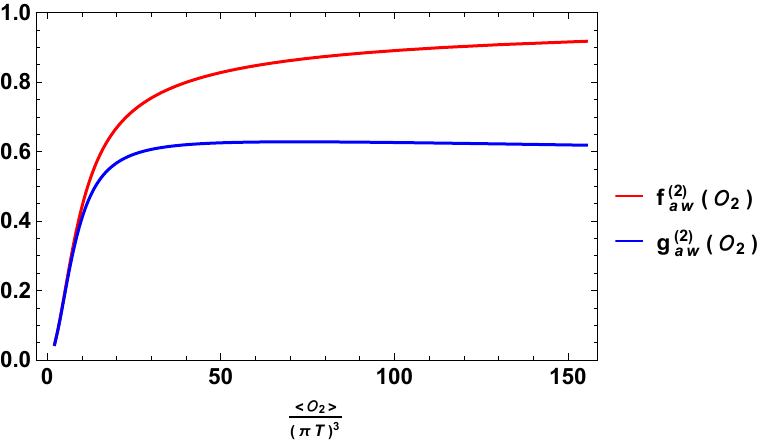}
\caption{{\bf Results for the $O_2$ theory.} Plots of the functions $f^{(2)}_{s}(O_2)$ and $g^{(2)}_{s}(O_2)$, where $s \in \{ v, a, w\}$, $f^{(2)}_{a w}(O_2)$ and $g^{(2)}_{a w}(O_2)$, as a function of condensate $\langle O_2 \rangle/ (\pi T)^3$. Color code: Red -  $f^{(2)}_{v}(O_2)$, Blue - $g^{(2)}_{v}(O_2)$ (upper left panel);  Red -  $f^{(2)}_{a}(O_2)$, Blue - $g^{(2)}_{a}(O_2)$ (upper right panel): Red -  $f^{(2)}_{w}(O_2)$, Blue - $g^{(2)}_{w}(O_2)$ (lower left panel);  Red -  $f^{(2)}_{aw}(O_2)$, Blue - $g^{(2)}_{aw}(O_2)$ (lower right panel). See fig.~\ref{fig:fva1} for further details.}
 \label{fig:fva2}
\end{figure}


To verify the consistency of our numerical results, we will perform an alternative analysis of the conductivities by setting $\bar\mu_{+} = n\bar \mu_-$ $(n \in \mathbb R)$, while varying~$\bar\mu_-$. Under these conditions, Eqs.~(\ref{eq:fivv1}) and (\ref{eq:fiaa1}) reduce to
\begin{eqnarray}
\sigma^B_{ss}(O_i) &=& \frac{\mu_-(O_i)}{4 \pi^2} \times \left( n f^{(i)}_{s}(O_i)+ g^{(i)}_{s}(O_i) \right) \,, \qquad \left( s \in \{ v, a \} \right) \,. \label{eq:fivv1_v2} 
\end{eqnarray}
A key advantage of this configuration is that, upon normalization by $\mu_-$, the result becomes independent of $\mu_-$, leaving a sole dependence on $O_i$ through the monotonically decreasing functions~$f_s^{(i)}(O_i)$ and~$g_s^{(i)}(O_i)$. Consequently, a consistency check of the numerical computation can be performed by comparing the numerical curves directly with the constant asymptotic values predicted analytically from Eq.~(\ref{eq:fivv1_v2}) and table~\ref{tab:fg_limits}. For $\sigma^{B}_{vv}$ and $\sigma^{B}_{aa}$, one finds
\begin{equation} 
\frac{4\pi^2}{\mu_-} \sigma^{B}_{vv} \to \{ n+1, n \}\,, \qquad \frac{4\pi^2}{\mu_-} \sigma^B_{aa} \to \left\{ n+1, \frac{n}{4} \right\} \,, \qquad    (O_i \to \{0,\infty\}) \,.  \label{eq:sigma_vv_aa_num}
		\end{equation}
		Following a similar procedure for Eqs.~(\ref{eq:fiww1}) and (\ref{eq:fiaw1}), we obtain
\begin{equation}
\frac{4\pi^2}{\mu_-} \sigma^B_{ww} \to \left\{0, \frac{n}{4} \right\} \,, \qquad \frac{4\pi^2}{\mu_-} \sigma^B_{aw} \to \left\{0, \frac{n}{4} \right\} \,, \qquad (O_i \to \{0,\infty\}) \,. \label{eq:sigma_ww_num}
\end{equation}
                We show in figs.~\ref{fig:fluct3} and \ref{fig:fluct-mu} the behavior of the chiral magnetic conductivities $\sigma^B_{ss}$ $(s \in \{v, a, w \})$ and $\sigma^B_{aw}$  in the $O_1$ theory and the $O_2$ theory, respectively, for the present case with $n=2$. As shown in these figures, the numerical results are in good agreement with the analytical values from Eqs.~(\ref{eq:sigma_vv_aa_num})-(\ref{eq:sigma_ww_num}), providing a robust consistency check for our numerical calculations. Notably, this convergence is significant slower for $\sigma^B_{vv}$ within the $O_2$ theory. In the case of $\sigma^B_{aw}$, exploring larger condensate values would be required to achieve a more precise verification.~\footnote{While Eq.~(\ref{eq:sigma_ww_num}) indicates that both $(4\pi^2 / \mu_-) \times \sigma^B_{ww}$ and $(4\pi^2 / \mu_-) \times\sigma^B_{aw}$ should reach the same asymptotic limit for large condensate values, a minor discrepancy is visible for the latter in the bottom-right panels of figs.~\ref{fig:fluct3} and~\ref{fig:fluct-mu}.  Nevertheless, as shown in fig.~11 of Ref.~\cite{Tamang:2025ekj} -- where explicit symmetry breaking was included in the model -- values of $\langle O_1 \rangle /(\pi T) \gtrsim 10$ are required to observe a clear convergence of~$\sigma^B_{aw}$ to its asymptotic limit. A similar behavior is expected in the presence of SSB.}

		\begin{figure}[htbp]
			\centering
			\includegraphics[width=.43\textwidth]{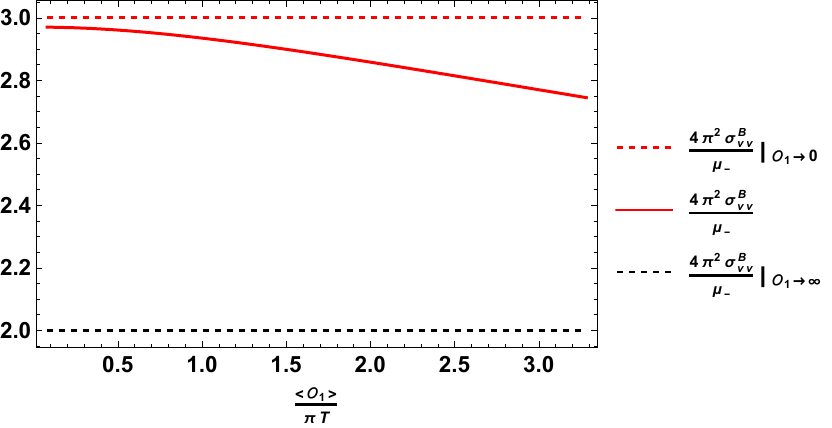}
			\includegraphics[width=.43\textwidth]{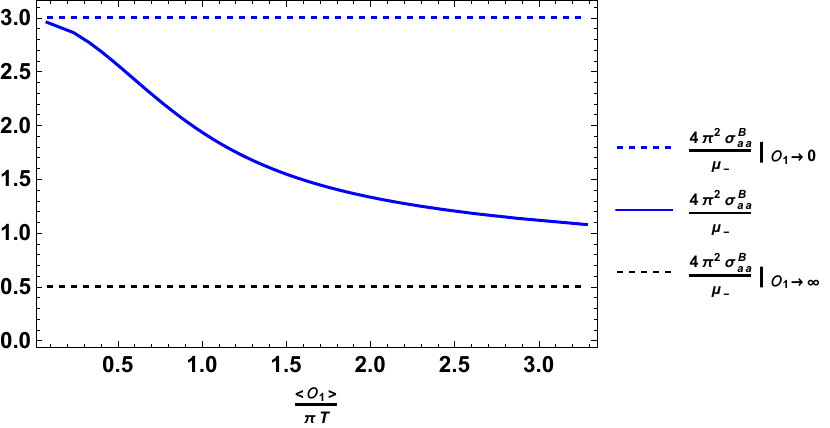}
			\includegraphics[width=.43\textwidth]{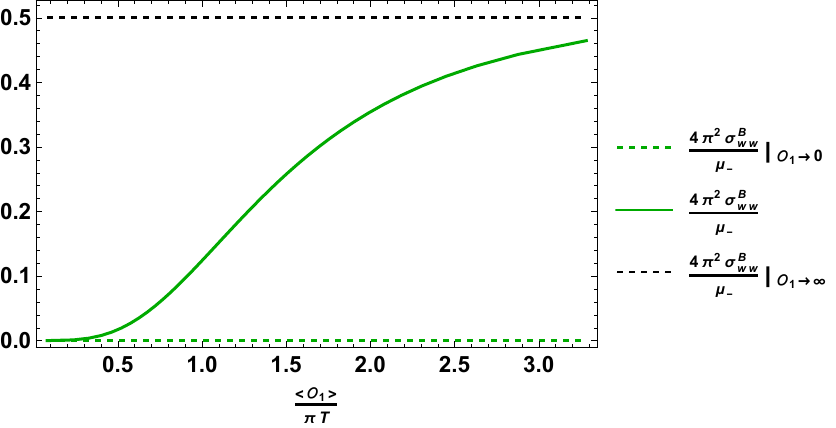}
			\includegraphics[width=.43\textwidth]{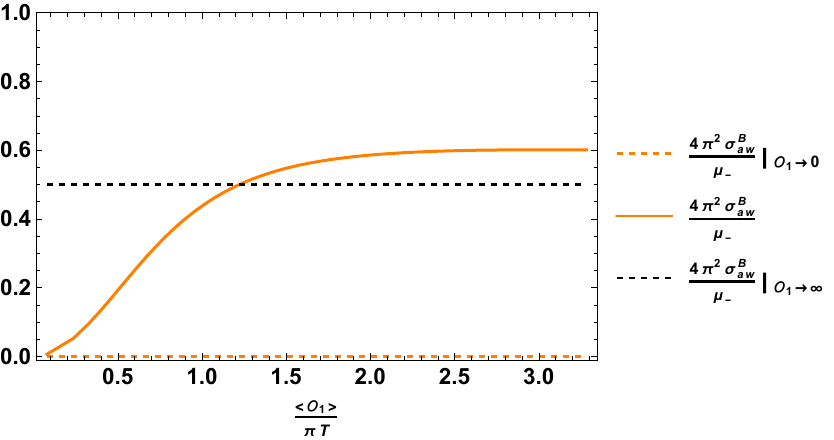}
			\caption{{\bf Results for the $O_1$ theory.} Plots of the conductivities $\sigma^{B}_{vv}$, $\sigma^{B}_{aa}$, $\sigma^{B}_{ww}$ and $\sigma^{B}_{aw}$, as a function of the condensate $\langle O_1 \rangle/ (\pi T)$. The dashed horizontal lines correspond to the analytic results given by Eqs.~(\ref{eq:sigma_vv_aa_num})-(\ref{eq:sigma_ww_num}): $\langle O_1 \rangle \to 0$ (shown in the same colors as the corresponding numerical curves) and $\langle O_1 \rangle / (\pi T) \to \infty$ (black). We have set $\bar\mu_{+} = 2\bar\mu_-$, while varying~$\bar\mu_-$.}
                        \label{fig:fluct3}
		\end{figure}

		\begin{figure}[htbp]
			\centering
			\includegraphics[width=.43\textwidth]{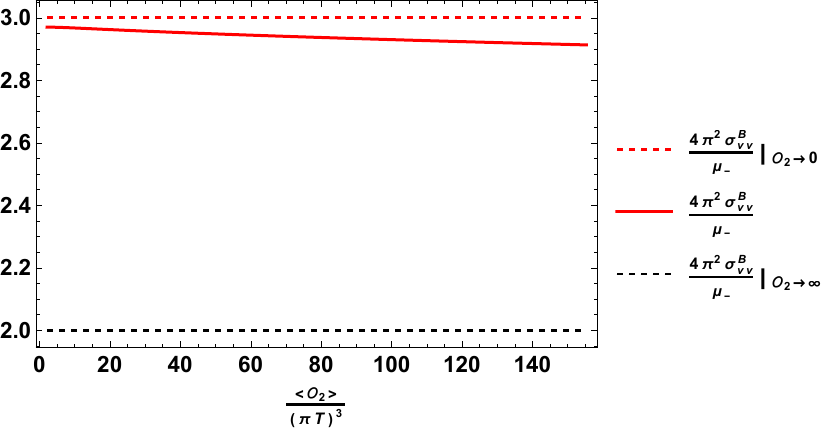}
			\includegraphics[width=.43\textwidth]{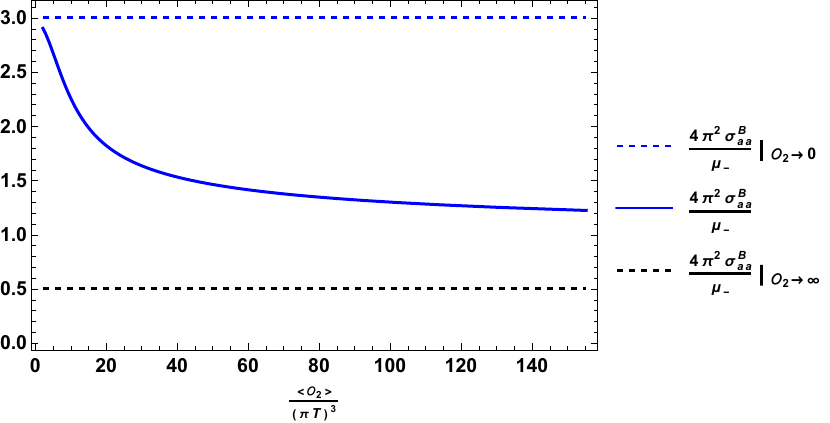}
			\includegraphics[width=.43\textwidth]{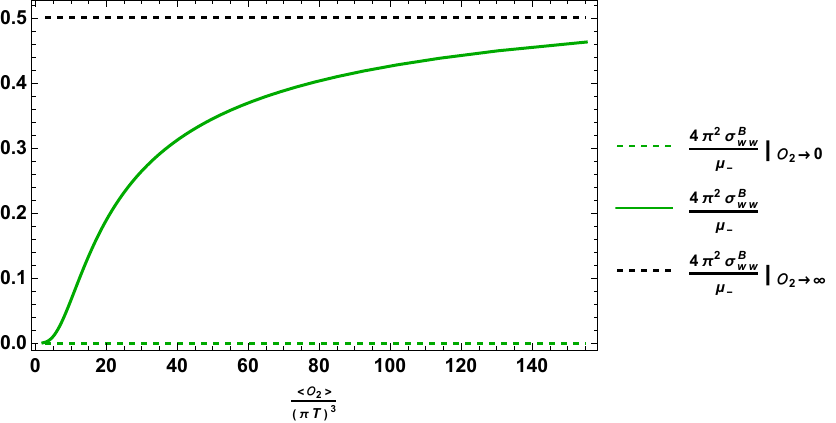}
			\includegraphics[width=.43\textwidth]{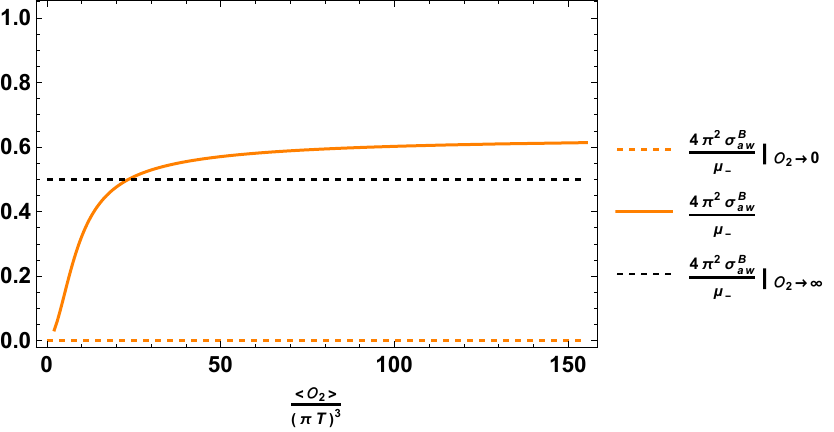}
			\caption{{\bf Results for the $O_2$ theory.} Plots of the conductivities $\sigma^{B}_{vv}$, $\sigma^{B}_{aa}$, $\sigma^{B}_{ww}$ and $\sigma^{B}_{aw}$, as a function of the condensate $\langle O_2 \rangle/ (\pi T)^3$. See fig.~\ref{fig:fluct3} for further details.}
 \label{fig:fluct-mu}
\end{figure}
                
         
\subsection{Results of the conductivities for zero chemical potentials}
\label{sec:result0}

For completeness and clarity, we present in figs.~\ref{fig:mv0mp01} and \ref{fig:mv0mp02}  the results of the non-vanishing conductivities for zero chemical potentials $(\bar\mu_v = \bar\mu_+ = 0)$ as a function of the condensates $O_1$ and $O_2$, respectively. The numerical results of the conductivities are compared with the results from the formulas proposed in section~\ref{subsec:disc_vv_aa_ww}, showing excellent agreement. The conductivities $\sigma^B_{av}=\sigma^B_{va}$ and $\sigma^B_{wv}=\sigma^B_{vw}$ are not displayed because they vanish identically in this regime, cf. Eq.~(\ref{eq:fivaw1}). Note that while we have fixed $\bar\mu_v=\bar\mu_+=0$, the chemical potential $\bar\mu_-$ cannot be set to zero in general as it is constrained by the values of the condensates.  The trend of the non-vanishing conductivities is similar to that observed in the finite chemical potential case: in particular, conductivities inducing $\vec{J}_a (\vec{J}_w)$ become suppressed(enhanced) as the condensates increase. 
\begin{figure}[htbp]
	\centering
	\includegraphics[width=.43\textwidth]{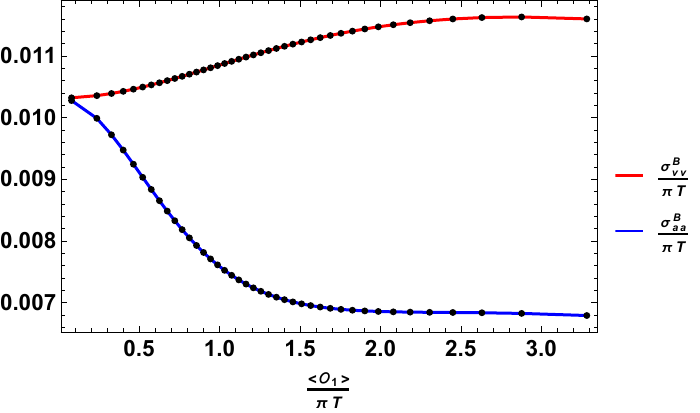}
	\includegraphics[width=.43\textwidth]{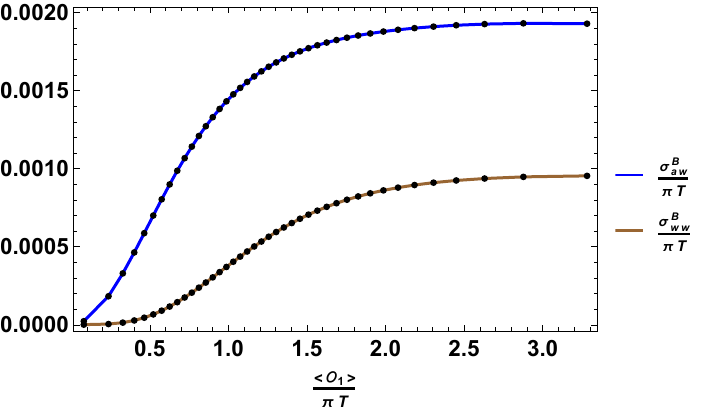}
	\caption{{\bf Results for the $O_1$ theory.} Plots of the conductivities $\sigma^{B}_{vv}$ and $\sigma^{B}_{aa}$ (left panel), and $\sigma^{B}_{ww}$ and $\sigma^{B}_{aw}=\sigma^{B}_{wa}$ (right panel), as a function of condensate $\langle O_1 \rangle/ (\pi T)$. The solid lines correspond to the original numerical results, while the black dots in each plot stand for the results obtained from Eqs.~(\ref{eq:fivv1})-(\ref{eq:fiaw1}) after imposing $ \bar\mu_+=0$. We have set $\bar\mu_v = \bar\mu_+=0$, while varying~$\bar\mu_-$.}
	\label{fig:mv0mp01}
\end{figure}

\begin{figure}[htbp]
	\centering
	\includegraphics[width=.43\textwidth]{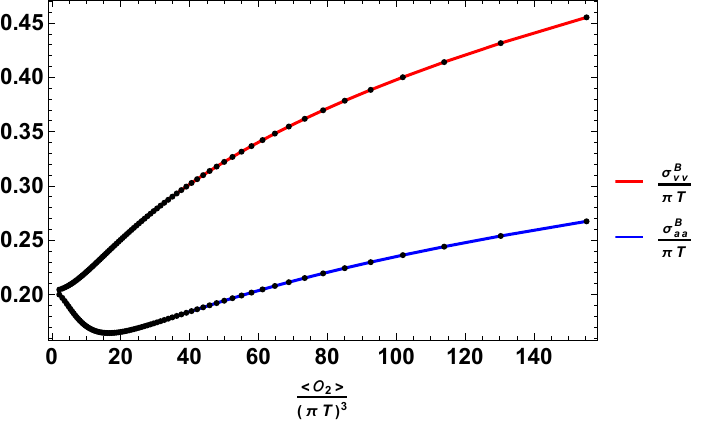}
	\includegraphics[width=.43\textwidth]{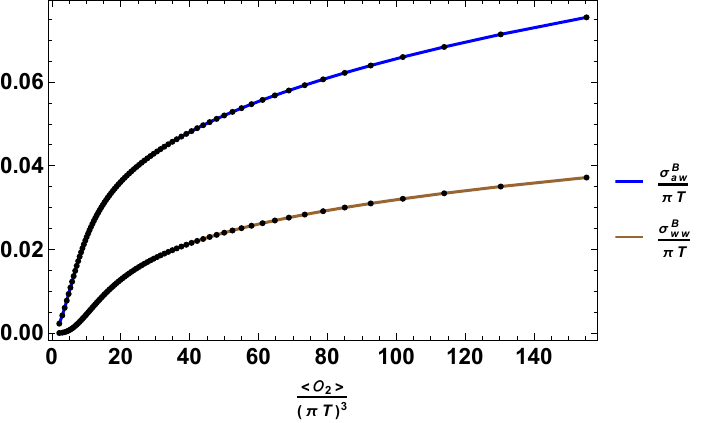}
	\caption{{\bf Results for the $O_2$ theory.} Plots of the conductivities $\sigma^{B}_{vv}$ and $\sigma^{B}_{aa}$ (left panel), and $\sigma^{B}_{ww}$ and $\sigma^{B}_{aw}=\sigma^{B}_{wa}$ (right panel), as a function of the condensate $\langle O_2 \rangle/ (\pi T)^3$. See fig.~\ref{fig:mv0mp01} for further details.}
	\label{fig:mv0mp02}
\end{figure}

\section{Discussion and outlook}
\label{sec:discussion}

In this work, we have studied the effect of spontaneous symmetry breaking on anomaly-induced transport by using a holographic framework that includes a pure gauge Chern-Simons term. The model includes vector and axial gauge fields, and an additional non-anomalous gauge field. The spontaneous symmetry breaking is introduced via a charged scalar field. After fixing the value of the scalar mass to $m^2=-3$, we have investigated two different quantizations giving rise to the $O_1$ theory and the $O_2$ theory, and corresponding to scalar dimensions $\Delta=1$ and $\Delta=3$, respectively. The numerical solutions of the background in both theories display a second‑order superconducting phase transition.

Our main result is that the effect of the anomaly is not restricted to the anomalous sector once the symmetry is spontaneously broken. In the normal phase, $O_i \to 0$ with $i=1,2$, all conductivities involving the non-anomalous $w$ current vanish, i.e.
\begin{equation}
	\sigma^B_{sw}=\sigma^B_{ws}=0\,,
	\qquad s\in\{v,a,w\}\,,
\end{equation}
whereas they become non-vanishing as the condensate develops. This behavior is observed in both the $O_1$ and the $O_2$ theories, showing that, within the present model, the effect of anomaly-induced transport on the non-anomalous sector is independent of the specific choice of quantization. We have characterized how anomalous conductivities depend on the condensate and chemical potentials. In particular, for $\bar{\mu}_v=\bar{\mu}_+=0$, the conductivities such as $\sigma^B_{vv}$, $\sigma^B_{aa}$, $\sigma^B_{ww}$, and $\sigma^B_{aw}=\sigma^B_{wa}$ remain non-zero in the presence of SSB, while $\sigma^B_{av}=\sigma^B_{va}$ and $\sigma^B_{wv}=\sigma^B_{vw}$ are vanishing. This choice allows us to study the effects of $\bar\mu_-$ and $O_{1,2}$ on the transport coefficients, by eliminating the contributions from $\bar\mu_v$ and $\bar\mu_+$.

There are several possible interesting extensions of the present study. First, going beyond the probe limit by including the backreaction of the matter fields on the geometry, and including the mixed gauge-gravitational Chern-Simons term, would allow one to study the chiral vortical effect and the energy transport sector, some features that are not captured by the present analysis. Second, extending this superconducting setup to a superfluid — by including non-zero spatial components of the gauge fields — could reveal how superfluid velocity influences anomaly-induced transport. Finally, it would be interesting to extend the analysis to finite frequencies and momenta in order to study the complete dynamical response and determine whether anomaly-induced transport in the non-anomalous sector persists beyond the static, long-wavelength limit.
   
\begin{acknowledgments}
  We would like to thank Karl Landsteiner for his collaboration on related work. A.T. would like to thank his supervisor Shubhrangshu Ghosh for his valuable suggestions and help throughout the work. E.M. would like to thank the Instituto de F\'{\i}sica of the Universidade de S\~ao Paulo, Brazil, and the ICTP South American Institute for Fundamental Research (SAIFR), S\~ao Paulo, Brazil, for their hospitality and financial support during the initial stages of this work. A.T. acknowledges financial support from the Ministry of Tribal Affairs, Government of India, through the National Fellowship for Scheduled Tribes (NFST). The work of N.R. and E.M. is supported by the ``Proyectos de Investigaci\'on Precompetitivos'' Program of the Plan Propio de Investigaci\'on of the University of Granada under grant PP2025PP-18.  The work of N.R. is also supported  by the Anusandhan National Research Foundation (ANRF), Government of India, under the Prime Minister Early Career Research Grant (PM ECRG) -  ANRF/ECRG/2025/001221/PMS.
		\end{acknowledgments}

		\begin{appendices}
			
			\section{Equations of motion of the fluctuations}
			\label{sec:appendix_EoM_fluctuations}
			
			We present in this appendix the explicit expressions of the equations of motion for the fluctuations. These are given by
\begin{flalign}
				0 &= v_x''
				+\frac{f'}{f}\,v_x' + i p \kappa \frac{2}{r_h^2 f}
				\Big(\mu_v \left((a_{-})_y
				+ (a_{+})_y\right)
				+\left(\mu_{+}-A_{t-}'\right)v_y\Big)  \,, \label{eq:vx} \\
				0 &= v_y''
				+\frac{f'}{f}\,v_y'
				- i p \kappa \frac{2}{r_h^2 f}
				\Big(
				\mu_v\left((a_{-})_x
				+ (a_{+})_x\right)
				+\left(\mu_{+}-A_{t-}'\right)v_x
				\Big) \,, \label{eq:vy} \\
				0 &= (a_{-})_x''
				+\frac{f'}{f}\,(a_{-})'_x
				-\frac{\phi^2}{u^2f}\,(a_{-})_x
				+i p \kappa \frac{1}{r_h^2 f}
				\Big(
				4\mu_v\,v_y
				+\left(\mu_{+}-A_{t-}'\right)
				\left((a_{-})_y
				+ (a_{+})_y\right)
				\Big) \,, \label{eq:a1x}  \\
				0 &= (a_{-})_y''
				+\frac{f'}{f}\,(a_{-})_y'
				-\frac{\phi^2}{u^2f}\,(a_{-})_y
				-ip\kappa \frac{1}{r_h^2 f}
				\Big(
				4\mu_v\,v_x
				+\left(\mu_{+}-A_{t-}'\right)
				\left((a_{-})_x+(a_{+})_x\right)
				\Big)\,, \label{eq:a1y} \\
				0 &= (a_{+})_x''
				+\frac{f'}{f}\,(a_{+})_x'
				+i p\kappa \frac{1}{r_h^2 f}
				\Big(
				4\mu_v\,v_y
				+\left(\mu_{+}-A_{t-}'\right)
				\left((a_{-})_y+(a_{+})_y\right)
				\Big) \,, \label{eq:a2x}  \\
				0 &= (a_{+})_y''
				+\frac{f'}{f}\,(a_{+})_y'
				-i p\kappa \frac{1}{r_h^2 f}
				\Big(
				4\mu_v\,v_x
				+\left(\mu_{+}-A_{t-}'\right)
				\left((a_{-})_x+(a_{+})_x\right)
				\Big)\,. \label{eq:a2y} 
			\end{flalign}
In these expressions, the spatial momentum is chosen along the $z$-axis, $\vec{p}=p\,\hat{z}$, and we have set $\omega=0$. Since the chiral magnetic conductivities are extracted from the terms linear in~$p$, we have retained the contributions to the equations of motion of the fluctuations up to $\mathcal{O}(p)$, and have neglected some non-anomalous contributions of~$\mathcal{O}(p^2)$.

\section{Analytical results of the conductivities}
\label{sec:sigma_analytic}

In the results presented in section~\ref{sec:result}, we observe the following properties for both zero and finite chemical potentials
\begin{equation}
\sigma^B_{sw}(O_i \to 0) = \sigma^B_{ws}(O_i \to 0) = 0 \,, \qquad (s \in \{v,a,w\}\,, \; i=1,2) \,.
  \end{equation}
This means that all conductivities related to the $w$ symmetry vanish in the zero condensate limit (i.e. the symmetric phase), as expected in the absence of SSB. The values of the remaining non-vanishing conductivities are given in Ref.~\cite{Tamang:2025ekj} (see table~\ref{tab:sigma_vaw_O0} below).

It is interesting to study the behavior of the conductivities in the large condensate limit. To derive analytical expressions in this regime, we will first explore the relation between the conductivities in two different bases for the gauge fields. Notably, the analytical results derived in this appendix correspond to fixed values of the chemical potentials $\mu_s$ $\left(s \in \left\{v, + , - \right\}\right)$. The dependence of $\mu_-$ on the condensates introduces certain deviations from these analytical expressions in some of the conductivities, a feature which is discussed in detail in section~\ref{subsec:disc_vv_aa_ww}.

		\subsection{Relation between bases}
		
		Let us study the relation between the conductivities in the two bases: $(v,+,-)$ and $(v,a,w)$. The gauge fields and currents in these bases are related by
		\begin{equation}
			A_\pm = A \pm W \,, \qquad J_\pm = \frac{1}{2} \left( J_a \pm J_w \right) \,,
		\end{equation}
		or, equivalently, the inverse relations
		\begin{eqnarray}
			&& A = \frac{1}{2} \left( A_+ + A_- \right) \,, \qquad W = \frac{1}{2} \left( A_+ - A_- \right) \,,  \\
			&& J_a = J_+ + J_- \,, \qquad\hspace{0.7cm} J_w =  J_+ - J_- \,.
		\end{eqnarray}
		By applying these identities, it is straightforward to obtain the following relation for the conductivity $\sigma^B_{av}$ in terms of the conductivities in the $(v,+,-)$ basis,
		\begin{equation}
			\sigma^B_{av} = \langle J_a J_v \rangle = \langle  (J_+ + J_-) \, J_v \rangle  =  \langle J_+  J_v \rangle +  \langle J_-  J_v \rangle  = \sigma^B_{+v} + \sigma^B_{-v}  \,.  \label{eq:sigmaav_2}
		\end{equation}
		By performing the same exercise for the other conductivities, one can easily obtain the results for the change of basis $(v,+,-) \rightarrow (v,a,w)$ given in table~\ref{tab:sigma_basis} (top panel). The results for the inverse operation, i.e. the change of basis $(v,a,w) \rightarrow (v,+,-)$, are also provided in table~\ref{tab:sigma_basis} (bottom panel).
	\begin{table}[htb]
			\centering
				\begin{tabular}{||c||c|c|c||}
					\hline
					$\sigma_{s s^\prime}^B$          & $v$  &  $a$   &  $w$    \\ 
					\hline \hline
					$v$  & $\sigma^B_{vv}$  & $\sigma^B_{v+} + \sigma^B_{v-}$ & $\sigma^B_{v+} - \sigma^B_{v-}$  \\
					\hline
					$a$  & $\sigma^B_{+v} + \sigma^B_{-v}$ & $ \sigma^B_{++} + 2 \sigma^B_{+-} + \sigma^B_{--} $ & $\sigma^B_{++} - \sigma^B_{--}$  \\
					\hline
					$w$  & $\sigma^B_{+v} - \sigma^B_{-v}$  & $ \sigma^B_{++} - \sigma^B_{--}$  & $\sigma^B_{++} - 2 \sigma^B_{+-} + \sigma^B_{--}$   \\
					\hline
    \midrule
                                        
                                      \hline
					$\sigma_{s s^\prime}^B$          & $v$  &  $+$   &  $-$    \\ 
					\hline \hline
					$v$  & $\sigma^B_{vv}$  & $\frac{1}{2} \left( \sigma^B_{va} + \sigma^B_{vw} \right)$ & $\frac{1}{2} \left( \sigma^B_{va} - \sigma^B_{vw} \right)$  \\
					\hline
					$+$  & $\frac{1}{2} \left( \sigma^B_{av} + \sigma^B_{wv} \right)$ & $\frac{1}{4} \left( \sigma^B_{aa} + 2 \sigma^B_{aw} + \sigma^B_{ww} \right)$ & $\frac{1}{4} \left( \sigma^B_{aa} - \sigma^B_{ww} \right)$  \\
					\hline
					$-$  & $\frac{1}{2} \left( \sigma^B_{av} - \sigma^B_{wv} \right)$  & $\frac{1}{4} \left( \sigma^B_{aa} - \sigma^B_{ww} \right)$  & $\frac{1}{4} \left( \sigma^B_{aa} - 2 \sigma^B_{aw} + \sigma^B_{ww} \right)$   \\
					\hline
				\end{tabular}
			\caption{\it Top [bottom] panel: Expressions for the chiral magnetic conductivities in the $(v,a,w)$ $[(v,+.-)]$ basis as linear combinations of the conductivities in the $(v,+,-)$ $[(v,a,w)]$ basis. It has been used in some cases the symmetry of the correlators: $\sigma^B_{+-} = \sigma^B_{-+}$ and $\sigma^B_{aw} = \sigma^B_{wa}$.}
			\label{tab:sigma_basis}
		\end{table}

\subsection{Conductivities in the $O_{1,2} \to 0$ and $O_{1,2} \to \infty$ limits}

		In the large $O_{1,2}$ limit, $A_-$ will be badly broken, and the chiral magnetic effects involving the $J_-$ current go to zero.  By using Eq.~(\ref{eq:sigmaav_2}), this implies that $\sigma^B_{av}(O_i \to \infty) = \sigma^B_{+v}(O_i \to \infty)$. Similar identities can be obtained for the other conductivities, and these are collected in table~\ref{tab:sigma_basis_largeO12}.
                \begin{table}[h!]
                  \centering
                                    \hspace{0.0cm}
                  \begin{minipage}{0.45\textwidth}
                    \centering
				\begin{tabular}{||c||c|c|c||}
					\hline
					$\sigma_{s s^\prime}^B(O_i \to \infty)$          & $v$  &  $a$   &  $w$    \\ 
					\hline \hline
					$v$  & $\sigma^B_{vv}$  & $\sigma^B_{v+} $ & $\sigma^B_{v+}$  \\
					\hline
					$a$  & $\sigma^B_{+v}$ & $\sigma^B_{++}$ & $\sigma^B_{++} $  \\
					\hline
					$w$  & $\sigma^B_{+v}$  & $\sigma^B_{++}$  & $\sigma^B_{++}$   \\
					\hline
				\end{tabular}
                  \end{minipage}
                  \hspace{0cm}
                  \begin{minipage}{0.45\textwidth}
                    \centering
				\begin{tabular}{||c||c|c|c||}
					\hline
					$\sigma_{s s^\prime}^B(O_i \to \infty)$          & $v$  &  $+$   &  $-$    \\ 
					\hline \hline
					$v$  & $\sigma^B_{vv}$  & $\sigma^B_{v+} $ & $0$  \\
					\hline
					$+$  & $\sigma^B_{+v}$ & $\sigma^B_{++}$ & $0$  \\
					\hline
					$-$  & $0$  & $0$  & $0$   \\
					\hline
				\end{tabular}
                  \end{minipage}
 \caption{\it Expressions for the chiral magnetic conductivities in the $O_i \to \infty$ limit $(i=1,2)$. The results are shown in the two bases: $(v,a,w)$ (left panel), and $(v,+,-)$ (right panel). It should be understood that all the conductivities in this table are evaluated at $O_i \to \infty$.}
			\label{tab:sigma_basis_largeO12}
                 \end{table}

In order to evaluate the asymptotic values of the conductivities $\sigma^B_{s s^\prime}$ in the limit $O_{1,2} \to \infty$, let us write the action of the model in the $(v, + , -)$ basis. The action is given by
		\begin{eqnarray}
		  \mathcal S^J &=& \frac{1}{16\pi G} \int d^5x \sqrt{-g} \bigg[ R + 2 \Lambda - \frac{1}{4} F_V^2 - \frac{1}{4} F_A^2 - \frac{1}{4} F_W^2  - J_v V - J_a A - J_w W\nonumber \\
                   &&\hspace{3cm} + \frac{\kappa}{6} A_- \wedge F_A \wedge F_A + \frac{\kappa}{6} A_+ \wedge F_A \wedge F_A  \nonumber \\
			&&\hspace{3cm} + \frac{\kappa}{2} A_- \wedge  F_V \wedge F_V  + \frac{\kappa}{2} A_+ \wedge  F_V \wedge F_V \bigg] + \mathcal S_{\textrm{GH}} \,, \label{eq:S_pm_v1}
		\end{eqnarray}
		where
		\begin{eqnarray}
                  F_A^2 &=& \frac{1}{4} \left( F_{A_+}^2 + F_{A_-}^2 + 2 F_{A_+} F_{A_-} \right) \,, \\
                 F_W^2 &=& \frac{1}{4} \left( F_{A_+}^2 + F_{A_-}^2 - 2 F_{A_+} F_{A_-} \right) \,, \\ 
			F_A \wedge F_A &=& \frac{1}{4}  F_{A_-} \wedge F_{A_-} +  \frac{1}{4} F_{A_+} \wedge F_{A_+} + \frac{1}{2} F_{A_-} \wedge F_{A_+}  \,.
		\end{eqnarray}
                We have included in Eq.~(\ref{eq:S_pm_v1}) the currents. Since $A_{-}$ is badly broken in the limit $O_{1,2} \to \infty$, one ends up with the following contributions to the action
		\begin{eqnarray}
		  \mathcal S^J &=& \frac{1}{16\pi G} \int d^5x \sqrt{-g} \bigg[ R + 2 \Lambda - \frac{1}{4} F_V^2 - \frac{1}{8} F_{A_+}^2 - J_v V - J_{+} A_+ \\
                    &&\hspace{3cm}  + \frac{\kappa}{24} A_+ \wedge F_{A_+} \wedge F_{A_+} + \frac{\kappa}{2} A_+ \wedge  F_V \wedge F_V  \bigg] + \mathcal S_{\textrm{GH}} \,. \nonumber
		\end{eqnarray}
		From a comparison with Eq.~(\ref{act}), one can see that both actions are related by the identifications of gauge fields and currents written in table~\ref{tab:S0_Sinfty}.
\begin{table}[htb]
\centering
\resizebox{5cm}{!}{
\begin{tabular}{||c|c||}
\hline
 $\mathcal S^J(O_{1,2} = 0)$  &  $\mathcal S^J(O_{1,2} \to +\infty)$    \\
        \hline\hline
   $\{V, J_v \}$ & $\{V, J_v \}$ \\
	\hline
    $\{ A, J_a \}$ & $\{A_+/2, 2J_+ \}$ \\
	  \hline
    $\{W, J_w\}$ & $\{A_+/2, 0\}$ \\
	\hline
\end{tabular} 
}
\caption{\it Gauge field and current identifications that allow to relate the action of the model in both limits: i) $O_{1,2} = 0$ (left column), and ii) $O_{1,2} \to +\infty$ (right column).}
\label{tab:S0_Sinfty}
\end{table}

Based on this result, the conductivities in the $O_{1,2} \to \infty$ limit can be derived from those of the symmetric theory ($O_{1,2} = 0$, cf. table~\ref{tab:sigma_vaw_O0}) by applying the following replacements
\begin{equation}
  \mu_v \to \mu_v \,, \qquad \mu_a \to \frac{\mu_+ }{2} \,, \qquad \mu_w \to \frac{\mu_+}{2} \,, \qquad \label{eq:transformation}
\end{equation}
along with the substitution  $J_a \to 2 J_+$ in the corresponding retarded correlators. Applying the transformations from Eq. (\ref{eq:transformation}) to the expressions in the left panel of table~\ref{tab:sigma_vaw_O0} yields the results presented in the right panel of table~\ref{tab:sigma_vaw_Oinf}. The remaining panels in both tables are obtained by a simple change of basis.

\begin{table}[h!]
  \centering
  \begin{minipage}{0.45\textwidth}
    \centering
\begin{tabular}{||c||c|c|c||}
					\hline
					$\sigma_{s s^\prime}^B(O_i \to 0)$          & $v$  &  $a$   &  $w$    \\ 
					\hline \hline
					$v$  & $\frac{\mu_a}{2\pi^2}$ & $\frac{\mu_v}{2\pi^2}$ & $0$  \\
					\hline
					$a$  & $\frac{\mu_v}{2\pi^2}$ & $\frac{\mu_a}{2\pi^2}$ & $0$  \\
					\hline
					$w$  & $0$  & $0$  & $0$   \\
					\hline
				\end{tabular} 
  \end{minipage}
  \hspace{0cm}
  \begin{minipage}{0.45\textwidth}
    \centering
\begin{tabular}{||c||c|c|c||}
					\hline
					$\sigma_{s s^\prime}^B(O_i \to 0)$          & $v$  &  $+$   &  $-$    \\ 
					\hline \hline
					$v$  & $\frac{\mu_+ + \mu_-}{4\pi^2}$ & $\frac{\mu_v}{4\pi^2}$ & $\frac{\mu_v}{4\pi^2}$  \\
					\hline
					$+$  & $\frac{\mu_v}{4\pi^2}$ & $\frac{\mu_+ + \mu_-}{16\pi^2}$ & $\frac{\mu_+ + \mu_-}{16\pi^2}$  \\
					\hline
					$-$  & $\frac{\mu_v}{4\pi^2}$  & $\frac{\mu_+ + \mu_-}{16\pi^2}$  & $\frac{\mu_+ + \mu_-}{16\pi^2}$   \\
					\hline
				\end{tabular} 
  \end{minipage}
  \caption{\it Chiral magnetic conductivities in the $O_{i} \to 0$ limit $(i=1,2)$. The results are shown in the two bases: $(v,a,w)$ (left panel), and $(v,+,-)$ (right panel).}
  \label{tab:sigma_vaw_O0}
 \end{table}
               

\begin{table}[h!]
  \centering
  \begin{minipage}{0.45\textwidth}
    \centering
\begin{tabular}{||c||c|c|c||}
					\hline
					$\sigma_{s s^\prime}^B(O_i \to \infty)$          & $v$  &  $a$   &  $w$    \\ 
					\hline \hline
					$v$  & $\frac{\mu_a + \mu_w}{4\pi^2}$ & $\frac{\mu_v}{4\pi^2}$ & $\frac{\mu_v}{4\pi^2}$  \\
					\hline
					$a$  & $\frac{\mu_v}{4\pi^2}$ & $\frac{\mu_a + \mu_w}{16\pi^2}$ & $\frac{\mu_a + \mu_w}{16 \pi^2}$  \\
					\hline
					$w$  & $\frac{\mu_v}{4\pi^2}$  & $\frac{\mu_a + \mu_w}{16 \pi^2}$  & $\frac{\mu_a + \mu_w}{16 \pi^2}$   \\
					\hline
\end{tabular} 
  \end{minipage}
  \hspace{0.0cm}
  \begin{minipage}{0.45\textwidth}
    \centering
\begin{tabular}{||c||c|c|c||}
					\hline
					$\sigma_{s s^\prime}^B(O_i \to \infty)$          & $v$  &  $+$   &  $-$    \\ 
					\hline \hline
					$v$  & $\frac{\mu_+}{4\pi^2}$ & $\frac{\mu_v}{4\pi^2}$ & $0$  \\
					\hline
					$+$  & $\frac{\mu_v}{4\pi^2}$ & $\frac{\mu_+}{16\pi^2}$ & $0$  \\
					\hline
					$-$  & $0$  & $0$  & $0$   \\
					\hline
\end{tabular} 
  \end{minipage}
  \caption{\it Chiral magnetic conductivities in the $O_{i} \to \infty$ limit $(i=1,2)$. The results are shown in the two bases: $(v,a,w)$ (left panel), and $(v,+,-)$ (right panel).}
 \label{tab:sigma_vaw_Oinf}
  \end{table}

Finally, table~\ref{tab:sigma_analytic} summarizes the analytical results for the conductivities in both limits: $O_i \to 0$ and $O_i \to \infty$ (cf. Ref.~\cite{Tamang:2025ekj}). As mentioned at the beginning of this appendix, an important caveat is that these expressions are obtained by fixing the values of the chemical potentials $\mu_v$, $\mu_+$ and~$\mu_-$. However, because $\mu_-$ grows with the condensate, conductivities with an explicit dependence on this chemical potential are expected to deviate from these predictions in the large condensate limit. As discussed in section~\ref{subsec:disc_vv_aa_ww}, the conductivities $\sigma^B_{va} = \sigma^B_{av}$ and $\sigma^B_{vw} = \sigma^B_{wv}$ agree remarkably well with the results in table~\ref{tab:sigma_analytic}, whereas the remaining conductivities are affected by these deviations. Inspired by the analytical results in this table, we have proposed in section~\ref{subsec:disc_vv_aa_ww} generic functional forms for the conductivities that successfully parameterize all of them.

		\begin{table}[htb]
			\centering
			\resizebox{10cm}{!}{
				\begin{tabular}{||c||c|c|c||}
					\hline
					$\sigma_{s s^\prime}^B(O_i= \{ 0, \infty \})$          & $v$  &  $a$   &  $w$    \\ 
					\hline \hline
					$v$  & $\left\{ \frac{\mu_+ + \mu_-}{4\pi^2} , \frac{\mu_+}{4\pi^2} \right\}$  & $\left\{ \frac{\mu_v}{2\pi^2} , \frac{\mu_v}{4\pi^2} \right\}$ & $\left\{ 0, \frac{\mu_v}{4\pi^2} \right\}$  \\
					\hline
					$a$  & $\left\{ \frac{\mu_v}{2\pi^2} , \frac{\mu_v}{4\pi^2} \right\}$ & $\left\{ \frac{\mu_+ + \mu_-}{4\pi^2} , \frac{\mu_+}{16\pi^2} \right\}$ & $\left\{0, \frac{\mu_+}{16 \pi^2} \right\}$  \\
					\hline
					$w$  & $\left\{ 0, \frac{\mu_v}{4\pi^2} \right\}$  & $\left\{0, \frac{\mu_+}{16 \pi^2} \right\}$  & $\left\{0, \frac{\mu_+}{16 \pi^2} \right\}$   \\
					\hline
				\end{tabular} 
			}
			\caption{\it Chiral magnetic conductivities in the symmetric theory, and in the large condensate limit. The values of the conductivities for $O_i = 0$ and $O_i \to \infty$ $(i=1,2)$ correspond to the left- and right-hand-side expressions in the curly brackets, respectively.}
			\label{tab:sigma_analytic}
		\end{table}

\section{Additional results for the conductivities}
\label{sec:appendix_sigma}

In this appendix, we present the numerical results for the chiral magnetic conductivities $\sigma^B_{s s^\prime}$ $\left(s,s^\prime \in \{v,a,w\}\right)$ as a function of $T/T_c$ for both the $O_1$ and $O_2$ theories.  These results are compared with the asymptotic limit $O_i \to 0$, and in certain cases, with the $O_i \to \infty$ limit. We note that $\langle O_1 \rangle / (\pi T) = 0$ for $T/T_c = 1$, whereas $\langle O_1  \rangle / (\pi T) \gg 1$ for $T/T_c \ll 1$, with analogous expressions holding for the $O_2$ theory (cf. bottom-right panels of figs.~\ref{fig:O1_theory} and~\ref{fig:O2_theory}). For clarity, these results are identical to those in sections~\ref{sec:result1} and \ref{sec:result2}, but are presented here as a function of $T/T_c$ rather than the condensates to provide an alternative perspective.

\subsection{Results in the $O_1$ theory}

We show in figs.~\ref{fig:app1}, ~\ref{fig:app2} and ~\ref{fig:app3} the chiral magnetic conductivities in the $O_1$ theory as a function of $T/T_c$.  All the conductivities associated with the non-anomalous symmetry ($\sigma^B_{wv} = \sigma^B_{vw}$, $\sigma^B_{wa} = \sigma^B_{aw}$ and $\sigma^B_{ww}$) vanish at $T = T_c$, and they increase as the temperature decreases. In contrast, the other conductivities are non-zero at $T = T_c$: $\sigma^B_{aa}$ and $\sigma^B_{av} = \sigma^B_{va}$ decrease as temperature drops, while $\sigma^B_{vv}$ remains largely independent of temperature.
			\begin{figure}[htbp]
				\centering
				\includegraphics[width=.43\textwidth]{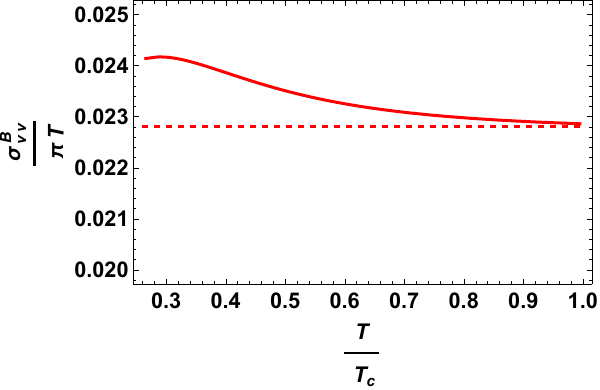}
				\includegraphics[width=.43\textwidth]{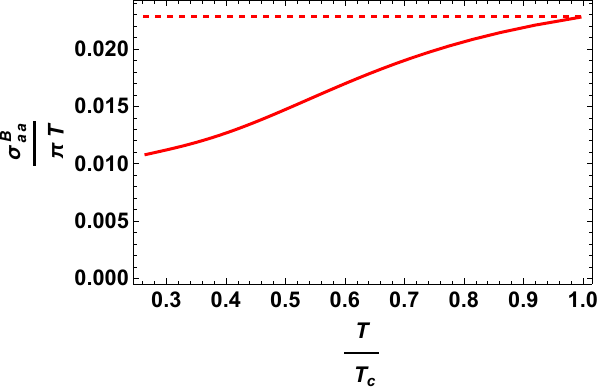}
				\caption{{\bf Results for the $O_1$ theory.} Plots of the conductivities $\sigma^{B}_{vv}$ (left panel) and $\sigma^{B}_{aa}$ (right panel), as a function of $T/T_c$. We have set $\bar\mu_v =0.1$ and $ \bar\mu_+=0.5$, while varying $\bar\mu_-$.  The dashed horizontal lines indicate the analytic limits: $\langle O_1 \rangle = 0$ at $T/T_c = 1$ (shown in the same colors as the corresponding numerical curves).}
        \label{fig:app1}
			\end{figure}
			
			\begin{figure}[htbp]
				\centering
				\includegraphics[width=.43\textwidth]{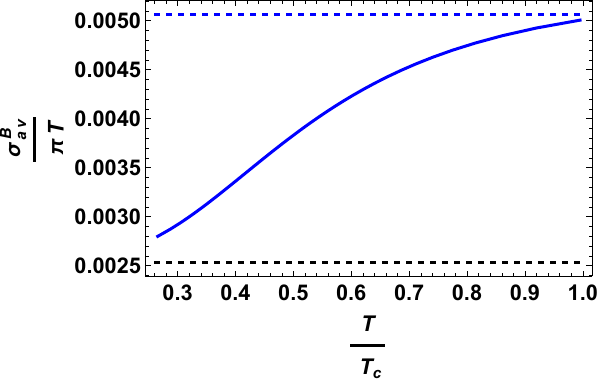}
				\includegraphics[width=.43\textwidth]{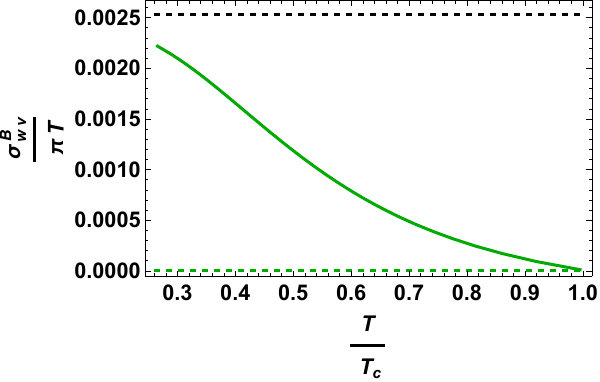}
				\caption{{\bf Results for the $O_1$ theory.} Plots of the conductivities $\sigma^{B}_{a v} = \sigma^{B}_{v a}$ (left panel) and $\sigma^{B}_{w v} = \sigma^{B}_{v w}$ (right panel), as a function of $T/T_c$. The dashed horizontal lines indicate the analytic limits: $\langle O_1 \rangle = 0$ at $T/T_c = 1$ (shown in the same colors as the corresponding numerical curves) and $\langle O_1 \rangle/(\pi T) \to \infty$ as $T / T_c \to 0$ (black). See fig.~\ref{fig:app1} for further details.}
  \label{fig:app2}
			\end{figure}
			
			\begin{figure}[htbp]
				\centering
				\includegraphics[width=.43\textwidth]{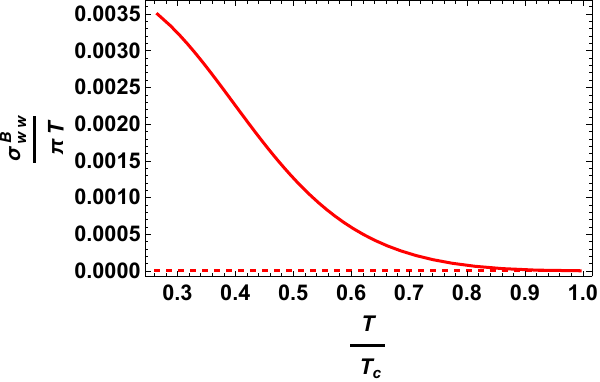}
				\includegraphics[width=.43\textwidth]{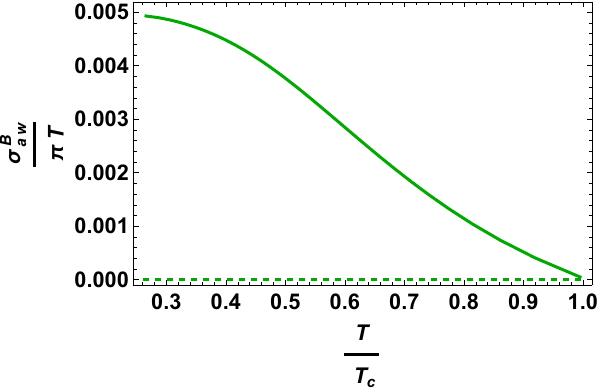}
				\caption{{\bf Results for the $O_1$ theory.} Plots of the conductivities $\sigma^{B}_{w w}$ (left panel) and $\sigma^{B}_{a v} = \sigma^{B}_{a w}$ (right panel), as a function of $T/T_c$. See fig.~\ref{fig:app1} for further details.}
                           \label{fig:app3}
			\end{figure}


\subsection{Results in the $O_2$ theory}

We show in figs.~\ref{fig:app4}, ~\ref{fig:app5} and ~\ref{fig:app6} the chiral magnetic conductivities in the $O_2$ theory as a function of $T/T_c$. The conclusions align with those for the $O_1$ theory; however, some differences appear in $\sigma^B_{vv}$ and $\sigma^B_{aa}$. Specifically, $\sigma^B_{vv}$ exhibits a stronger temperature dependence, whereas $\sigma^B_{aa}$ displays a non-monotonic behavior.
			\begin{figure}[htbp]
				\centering
				\includegraphics[width=.43\textwidth]{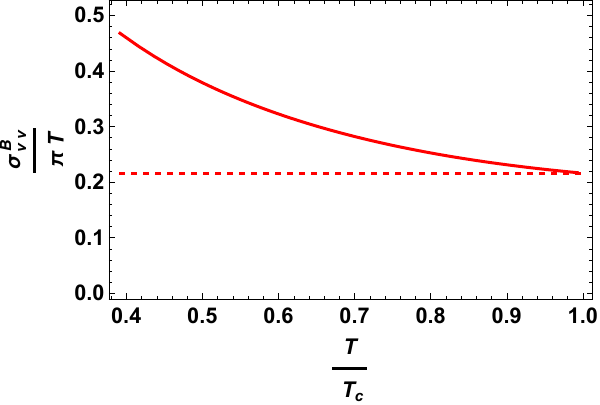}
				\includegraphics[width=.43\textwidth]{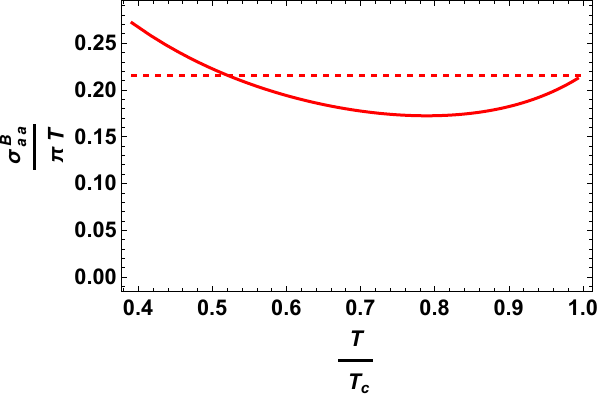}
				\caption{{\bf Results for the $O_2$ theory.} Plots of the conductivities $\sigma^{B}_{vv}$ (left panel) and $\sigma^{B}_{aa}$ (right panel), as a function of $T/T_c$.  See fig.~\ref{fig:app1} for further details.}
          \label{fig:app4}
			\end{figure}
			
			\begin{figure}[htbp]
				\centering
				\includegraphics[width=.43\textwidth]{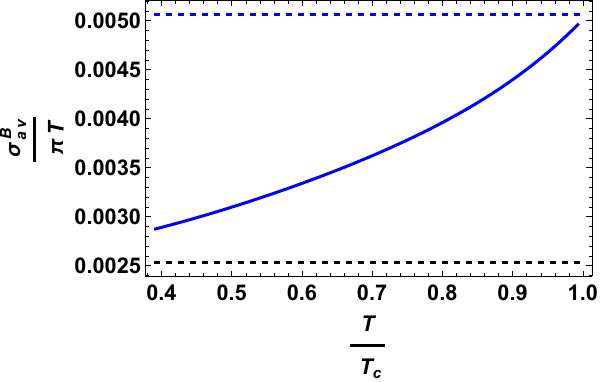}
				\includegraphics[width=.43\textwidth]{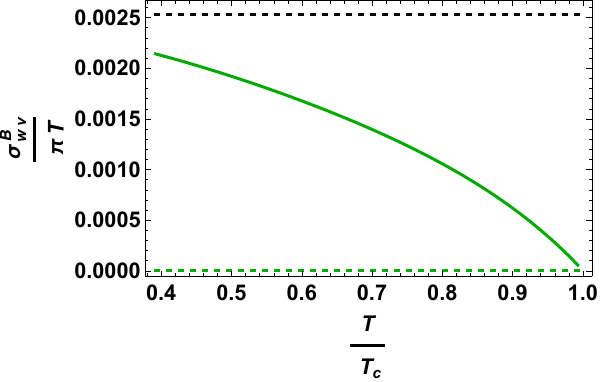}
				\caption{{\bf Results for the $O_2$ theory.} Plots of the conductivities $\sigma^{B}_{a v} = \sigma^{B}_{v a}$ (left panel) and $\sigma^{B}_{w v} = \sigma^{B}_{v w}$ (right panel), as a function of $T/T_c$. See figs.~\ref{fig:app1} and \ref{fig:app2} for further details.}
               \label{fig:app5}
			\end{figure}
			
			\begin{figure}[htbp]
				\centering
				\includegraphics[width=.43\textwidth]{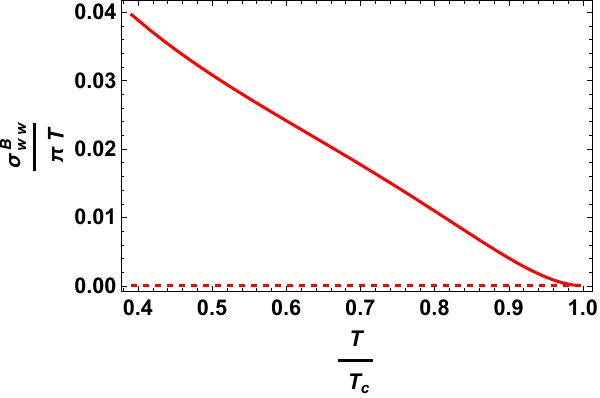}
				\includegraphics[width=.43\textwidth]{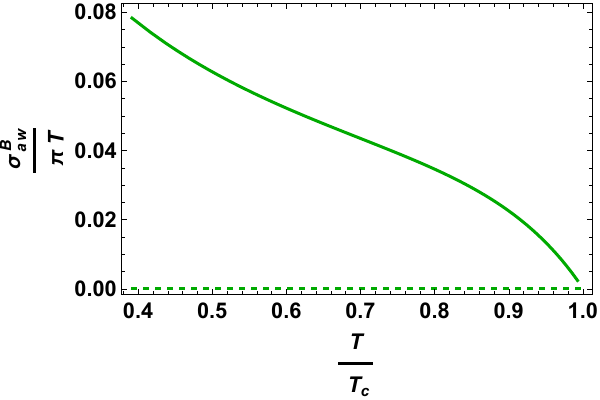}
				\caption{{\bf Results for the $O_2$ theory.} Plots of the conductivities $\sigma^{B}_{w w}$ (left panel) and $\sigma^{B}_{a v} = \sigma^{B}_{a w}$ (right panel), as a function of $T/T_c$. See fig.~\ref{fig:app1} for further details.}
          \label{fig:app6}
			\end{figure}

		\end{appendices}
		
		\bibliographystyle{unsrt}
		\bibliography{mybibnotes}

	\end{document}